\documentclass[namedreferences]{solarphysics}
\usepackage[hyperref,optionalrh,solaromanenum]{spr-sola-addons} 
\usepackage{graphicx}                    
\usepackage{color}                       
\usepackage{breakurl}                         

\begin{document}

\begin{article}

\begin{opening}

\title{Soft X-ray Pulsations in Solar Flares}

%
\author{P.~J.~A.~\surname{Sim{\~ o}es}$^{1}$\sep
        H.~S.~\surname{Hudson}$^{1,2}$\sep
        L.~\surname{Fletcher}$^{1}$      
       }

%
\runningauthor{Sim{\~ o}es et al.}
\runningtitle{GOES Pulsations}

%
  \institute{$^{1}$ SUPA, School of Physics and Astronomy, University of Glasgow, G12 8QQ, UK\\
                   email: \url{paulo.simoes@glasgow.ac.uk} email: \url{lyndsay.fletcher@glasgow.ac.uk}\\ 
           $^{2}$ SSL, UC Berkeley, CA 94720, USA
                     email: \url{hhudson@ssl.berkeley.edu} \\
             }

\begin{abstract}
The soft X-ray emissions ($h\nu > 1.5$~keV) of solar flares come mainly from the bright coronal loops at the highest temperatures normally achieved in the flare process.
Their ubiquity has led to their use as a standard measure of flare occurrence and energy, although the overwhelming bulk of the total flare energy goes elsewhere. 
Recently \citet[Astrophys. J. Lett. 749 L16]{2012ApJ...749L..16D} noted quasi-periodic pulsations (QPP) in the soft X-ray signature of the X-class flare SOL2011-02-15, as observed by the standard photometric data from the GOES (\textit{Geostationary Operational Environmental Satellite}) spacecraft.
In this paper we analyze the suitability of the GOES data for this kind of analysis and find them to be generally valuable after September, 2010 (GOES-15).
We then extend the Dolla \textit{et al.} result to a complete list of X-class flares from Cycle~24, and show that most of them (80\%) display QPPs in the impulsive phase.
The { pulsations} show up cleanly in both channels of the GOES data, making use of time-series of irradiance differences (the digital time derivative on the 2-s sampling). We deploy different techniques to characterize the periodicity of GOES pulsations, considering the red-noise properties of the flare signals,  { finding a range of chracteristic time scales of the QPPs for each event, but usually with no strong signature of a single period dominating in the power spectrum}.
The QPP may also appear on somewhat longer time scales during the later gradual phase, possibly with a greater tendency towards coherence, but the sampling noise in GOES difference data for large irradiance values (X-class flares) makes these more uncertain. We show that there is minimal phase difference between the differenced GOES energy channels, or between them and the hard X-ray variations on short time scales. During the impulsive phase the footpoints of the newly-forming flare loops may also contribute to the observed soft X-ray variations.
\end{abstract}

%

\end{opening}

%
 \section{Introduction}\label{sec:intro} 

The soft X-ray (SXR) emission of a solar flare (its ``GOES class'') has practically supplanted the H$\alpha$ brightening as the defining property of a solar flare \citep[\textit{e.g.}][]{2011SSRv..159...19F}. { The {\em Geostationary Operational Environmental Satellite} (GOES) is a satellite series initially deployed in 1974, 
with GOES-15 being the latest in the line. The GOES ``Sun-as-a-star'' soft X-ray observations, standardized in essentially the same form since the 1970s, come from two photometers that sample 1--8~\AA~and 0.5--4~\AA~bands.

We understand the soft X-ray sources to consist of plasma ablated (``evaporated'') from the lower solar atmosphere into coronal magnetic structures, and that the residence time scale (for cooling and draining) substantially exceeds the injection time scale.
The time of injection corresponds to the ``impulsive'' phase of the flare, although it can be prolonged in some cases to tens of minutes; the subsequent gradual phase sees continued X-ray emission for many hours in extreme cases.

The 1--8 and 0.5--4 \AA\ bands reflect a complicated mixture of line and continuum contributions \citep[\textit{e.g.}][]{2005SoPh..227..231W} and respond to temperatures above about 4~MK.}
The temperature sensitivity of this filter pair strongly favors flare emissions but there is sufficient sensitivity in the long-wavelength band to detect quiescent active regions as well, at least during active times.
The two broad passbands overlap the thermal/non-thermal regions of solar X-ray emission, and generally  reflect bound-bound transitions at lower temperatures, and continuum components (thermal free-free and free-bound) at higher temperatures.
Although the spectral response extends above 10~keV, the GOES time histories of solar flares typically do not show an impulsive-phase signature, and invariably lag the hard X-ray peak emission time, a manifestation of the \cite{1968ApJ...153L..59N} effect.
We attribute this to the rapid filling of the coronal magnetic reservoir with hot plasma, which gradually obscures the hot footpoint sources detectable in soft X-ray images \citep[][in prep.]{1993ApJ...416L..91M,1994ApJ...422L..25H,2004A&A...415..377M,2015Simoes}.
By comparison with the extreme ultraviolet (EUV) passbands \cite[\textit{e.g.}][]{2010A&A...521A..21O}, the GOES photometry isolates the hottest part of the coronal plasma.

Pulsations in the X-ray emission of solar flares were discovered in hard X-rays (above 20~keV) by \cite{1969ApJ...155L.117P} in  the impulsive phase of a major flare (SOL1968-08-08T18), and had a timescale of about 16~s. Simultaneous and closely synchronous peaks appeared at 15.4~GHz, identifiable as gyrosynchrotron radiation from MeV-range electrons, and \cite{1969ApJ...158L.127J} noted earlier pulsations at longer radio wavelengths with somewhat longer (24~s) spacings.
These longer-wavelength microwaves normally would be interpreted as thermal free-free radiation, and so in this early example we already could see pulsation signatures in very different physical domains.
In our current interpretation of flare hard X-ray emission, this behavior reflects time variations in an ill-understood mechanism for particle acceleration, by which the flare magnetic energy release goes predominantly into fast electrons at energies above about 20~keV.

{ The term ``quasi-periodic pulsations'' (QPP) is usually employed to refer generally to this kind of behavior -- short trains of roughly periodic emissions. Throughout the literature, the QPPs are usually identified by periodic signatures in the Fourier or wavelet power spectrum of the flare signal}. 

Since those early observations, a great deal of progress has been made in ``coronal seismology,'' in which one attempts to identify the QPP with resonant MHD wave properties and thereby learn something about the physical conditions in the corona \citep[\textit{e.g.}][]{2005LRSP....2....3N}.
{\cite{2009SSRv..149..119N} specifically review the now extensive observations of flare-related QPP mainly in this context.}
The identification of discrete eigenmodes during the early stages of a solar flare may not make much sense, though, since the flaring structure must evolve rapidly during this time and thus substantially change its geometry until the energy release subsides. 
On the other hand, structures external to the flare site clearly react to the disturbance in an oscillatory fashion \citep{1999ApJ...520..880A}, which may also involve a relaxation to a new equilibrium state \citep[\textit{e.g.}][]{2013ApJ...777..152S}.
Far-reaching disturbances have long been known to emanate from flares \cite[\textit{e.g.}][]{1976sofl.book.....S} and can frequently be identified with the shock-wave exciters of type~II radio bursts \citep{2004ApJ...614L..85H}.
Nowadays the EUV observations from the TRACE and SDO satellites in particular have provided a wealth of new observations via high-resolution movies of coronal structures.

Soft X-ray data (and EUV imaging at wavelengths corresponding to higher temperatures) in principle probe a frontier area between the dynamical process of energy release, and the perhaps oscillatory relaxation into the new equilibrium of the flaring active region.
We note early observations of X-ray emission lines by the soft X-ray spectrometer on board the \textit{Yohkoh} satellite \citep{2005ApJ...620L..67M,2006ApJ...639..484M}, making use mainly of S~{\sc xv} and Ca~{\sc xix} lines.
These observations reveal ``average oscillation periods'' of $5.5 \pm 2.7$~min, shorter than the typical periods associated with the large-scale structures seen in the EUV.
The flare QPP phenomena discovered by  \cite{1969ApJ...155L.117P} typically have time scales more than one order of magnitude shorter, generally suggesting smaller scale lengths, stronger magnetic fields, and lower altitudes.

In an important recent comment, \cite{2011A&A...533A..61G} \citep[see also][]{Vaughan:2005,Vaughan:2010} pointed out that an underlying power-law distribution of variability could mimic the appearance of a resonant process for a short time-series as an artifact of band-limited noise.
In such a case the apparent discrete frequencies of power-spectrum analysis such as that of \cite{1969ApJ...155L.117P}, and many subsequent examples, could readily be misinterpreted \citep{2015ApJ...798....1I,InglisIrelandDominique:2015}.

In this paper we characterize time variability of the GOES soft X-ray fluxes, following the discovery by \cite{2012ApJ...749L..16D} of QPP in the 1--30~s period range in the first X-class flare of Cycle~24, the well-studied SOL2011-02-15. Such QPPs were also noted by \cite{2013ApJ...777..152S} in the M-class flare SOL2012-03-09, with a similar behaviour observed in SDO/AIA EUV ``hot'' channels.
We analyse the whole set of X-class flares from this cycle to date, a total of 35 at the time of writing, in an effort to understand the relationship between the relatively fast impulsive-phase QPP seen in the nonthermal signatures, with the longer periods with possible resonant properties seen in thermal signatures formed in coronal magnetic structures with relatively low plasma temperatures.

\section{GOES data analysis}\label{sec:goes}

\begin{figure}[htbp]
\centerline{\includegraphics[width=0.5\textwidth,clip=]{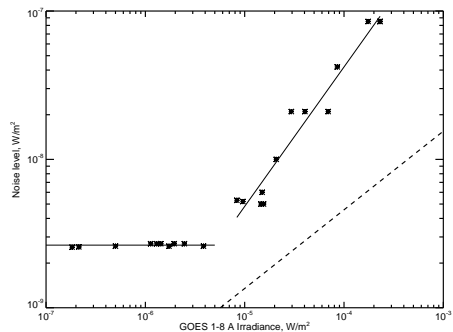}}
\caption{Empirical noise levels for GOES irradiance measurements in the $1--8$~\AA~band. The dashed line shows the estimated Poisson noise from photon counting statistics, and the points show the irradiance steps resulting from the digitization of the data.
We derived these levels from a series of flares covering the range of irradiance values shown in the plot.}
\label{fig:goes_noise}
\end{figure}

To sharpen the variability, we take advantage of the excellent signal-to-noise ratio of the { current} GOES soft X-ray irradiance data and
mainly study the { \textit{digital time-derivative}} of the original high-resolution (2~s {sampling}) time-series. 
{ The data prior to GOES-13 had poorer sampling and thus are not so useful for this kind of analysis.}
Since the time derivative of the SXRs is known to track the hard X-rays (HXR), peaks in the { derivative} time series should correspond well with HXR spikes, if both reflect the impulsive energy release (the Neupert effect).
However differencing data can degrade their precision, so we first study this issue.
A casual inspection of the GOES irradiance data shows discrete steps in the noise level imposed by the digitization, and Figure~\ref{fig:goes_noise}
characterizes this as a function of irradiance level for the 1--8~\AA~band.
We have taken observed irradiance digital steps from several flares, ranging from C~to X~class, in the time frame 2012--2014, and the solid lines
show a fit to the points.
We have estimated the Poisson noise from counting statistics and also show this, noting that it lies below the digital step size.
Because of this the precision of the GOES data does not increase with irradiance level, as should be the case for finer sampling  \citep[see also][]{2009SPIE.7438E..02C}.

\begin{figure}[htbp]
\centerline{\includegraphics[width=0.5\textwidth]{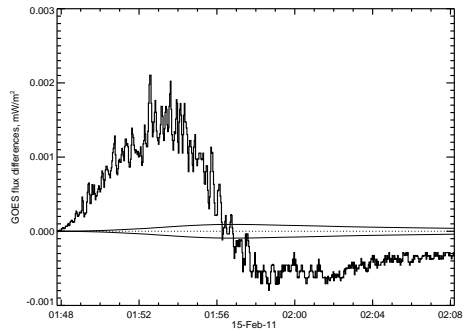}}
\caption{The { time derivative of} 1--8~\AA~GOES data for SOL2011-02-15, with the solid lines showing the digitization errors $\pm$ from the fit in Figure~\ref{fig:goes_noise}.}
\label{fig:SOL2011-02-15_show_goes}
\end{figure}

The iconic flare SOL2011-02-15 illustrates some of the properties of the differenced time-series; see Figure~\ref{fig:SOL2011-02-15_show_goes}.
For the purposes of this article we note a working definition of the two phases of this flare: we identify the impulsive phase as between GOES onset and the first zero crossing of the 1--8~\AA~differences, with the gradual phase the time between that zero crossing and the GOES end time.
In this flare the zero-crossing time is 01:56:49~UT, whereas the listed time of GOES maximum is 01:56~UT, and this distinction will normally not matter.
The strong variability seen in the impulsive phase of this flare has a rough time scale of about 12~s, estimated either by peak-counting or a power spectrum, which is similar to the original observation of hard X-ray pulsations (about 16~s) by Parks and Winckler.
The variability in the gradual phase is much less pronounced, and indeed has a magnitude generally comparable to the digitization
error (Figure~\ref{fig:goes_noise}).

We conclude from this representative event that the differencing technique properly reveals the QPP of the impulsive phase, but that the variability in the gradual phase competes in a complicated way with the digitization levels of the GOES data; we do not believe that this can be characterized quantitatively because the digital levels do not adequately sample the noise, meaning that successive points may have identical values reported.
This limitation should be borne in mind for the survey material presented below, which deals with X-class flares of comparable magnitudes to SOL2011-02-15.

 As further evidence of the fidelity of the QPP signature in the { time derivative of} GOES data for the impulsive phase, we show another example (Flare No. 4 of the survey; see below) in Figure~\ref{fig:SOL2011-09-07_twocolor}.
This compares the two independent GOES bands (0.5--4~\AA~and 1--8~\AA), whose ratio characterizes the spectral hardness.
The variations match well at the shorter time scales (the QPP), with some soft-hard-soft pattern indicated by the hardness ratio \cite[\textit{e.g.}][]{2011SSRv..159...19F} on the longer time scales as well.

\begin{figure}[htbp]
\centerline{\includegraphics[width=0.5\textwidth]{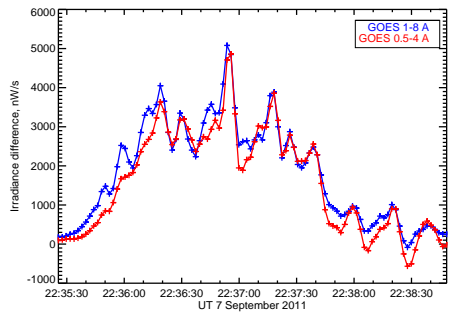}}
\caption{Correlation of { time-derivative of} GOES 1--8~\AA~ with GOES 0.5--4~\AA~ (scaled up by a factor 2.35) for SOL2011-09-07.}
\label{fig:SOL2011-09-07_twocolor}
\end{figure}

\section{Survey of X-class flares}

We have used the { time derivative of} GOES timeseries with all of the X-class flares in Cycle~24, up to the time of writing and commencing with Event~0 (SOL2011-02-15, the subject of the \cite{2012ApJ...749L..16D} paper and Figure~\ref{fig:SOL2011-02-15_show_goes}.
This sample contains 35 events, of which the strongest was No. 2, SOL2011-03-09 (X6.9).
The basic analysis tool is wavelet transforms, as described below (Section~\ref{sec:wavelet}), with additional checks.
All of these events showed QPP in the impulsive phase.

\subsection{Wavelet analysis}\label{sec:wavelet}

In order to characterise the variability in the SXR signal, we employ wavelet analysis using the methods and routines provided by \cite{TorrenceCompo:1998}. { To evaluate the red-noise spectrum, we have used the method described in their section 4, and provided by their wavelet software. The power-law slope that resulted from this was $-1.9$ for all the events in our sample.} As described, we use the { time derivative of} the GOES channels instead of the more usual approach of ``detrending'' the signal \citep[\textit{e.g.}][]{2012ApJ...749L..16D}, \textit{i.e.} subtracting a smoothed curve, which depends on an arbitrarily chosen time scale. 
The { time derivative} maintains the fluctuations in the signal, retaining the original periodicity (if present). For completeness, however, we have also applied the detrending approach, and found similar results. 
We investigated the presence of periodic signals in the local wavelet power spectrum (WPS), and in the WPS averaged over the impulsive and gradual phases of each event in our sample. 
The significance of the WPS is tested against a red-noise model, at 95\% { and 99.7\% confidence levels}; see \cite{2011A&A...533A..61G,InglisIrelandDominique:2015,2015ApJ...798....1I} for a discussion of this kind of analysis.
We used the Morlet mother wavelet (parameter $w=6$), but note that similar results were found for varying $w$ and also for the Paul mother wavelet \citep[see][for details]{TorrenceCompo:1998}. { In short, we define the presence of QPP as the identification of power in the wavelet power spectrum (averaged over the impulsive phase) above the red-noise model, with a confidence level of 99.7\%}.

\subsection{Survey results}

For our sample of 35 X-class flares, we have characterized the time variability in several basic ways, as summarized in Table~\ref{tab:bigtable}. 
For each event, we found the period range where the wavelet power spectrum, averaged for the impulsive and gradual phases, were above the 95\% confidence level of the red-noise model. 
We also verified the periods of peaks in the wavelet power spectrum. 

\begin{figure}[htbp]
\centerline{\includegraphics[angle=0,width=0.99\textwidth]{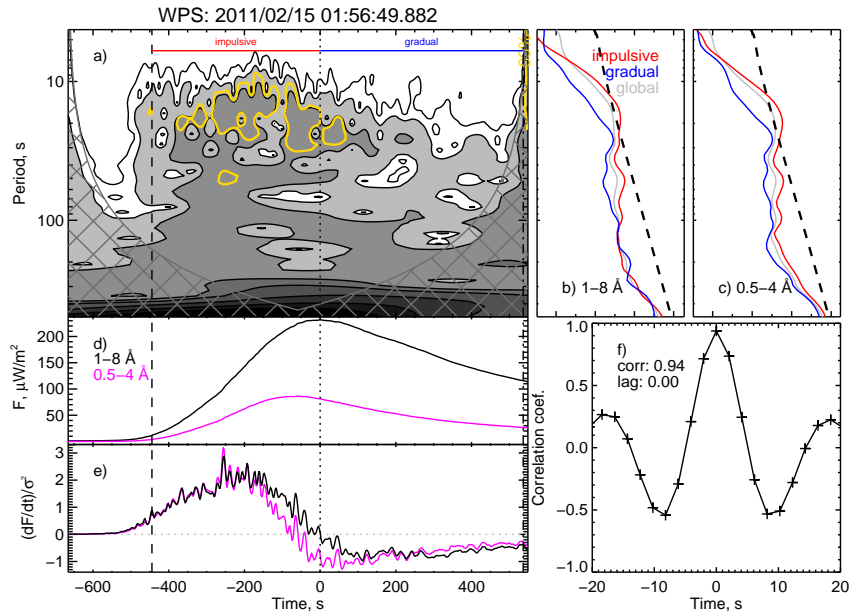}}
\caption{Results of wavelet analysis of SOL2011-02-15. Reference times are indicated by the vertical lines: onset (dashed), peak (dotted) and end (dash-dotted). a) wavelet power spectrum of the { time derivative of the} signal of the 1--8 \AA~ channel, with the power above the red-noise model (99.7\% confidence level) shown in yellow, b) the time-averaged power spectrum for the low channel (1--8 \AA) for the impulsive phase (red) and gradual phase (blue) compared to the 99.7\% confidence level of the red-noise spectrum model (thick dashed curve) and c) same for the high channel (0.5--4 \AA), d) timeseries of both GOES channels, e) { time derivative} of both channels, and f) lag cross-correlation between the { time derivative of the} signal of both channels.}
\label{fig:wavelet}
\end{figure}

As an example, in Figure \ref{fig:wavelet} we show the results of the wavelet analysis for our first event, SOL2011-02-15, also analysed by \cite{2012ApJ...749L..16D}. 
The four panels of Figure~\ref{fig:wavelet}  show  (a) the wavelet power spectrum, (b and c) the time-averaged power spectrum for the two GOES spectral bands, (d) the GOES timeseries, (e) the { time derivative of the} timeseries of both channels, and (f) the cross-correlation between the two channels, confirming the oscillatory {non-dispersive} nature of the signals (0.94 correlation coefficient with effectively zero lag). 
The wavelet power spectrum shows a periodic signal above the red-noise { 99.7\% confidence level} during the impulsive phase in the range 12--24 seconds, in agreement with the findings of \cite{2012ApJ...749L..16D}. 
In the gradual phase, the periodic power starts to vanish, but still appears to last for about 150 seconds, with a period range 20--40 seconds {(see Figure~\ref{fig:SOL2011-02-15_show_goes} for the caveat about sampling precision)}.

\begin{figure}[htbp]
\centerline{\includegraphics[width=0.9\textwidth]{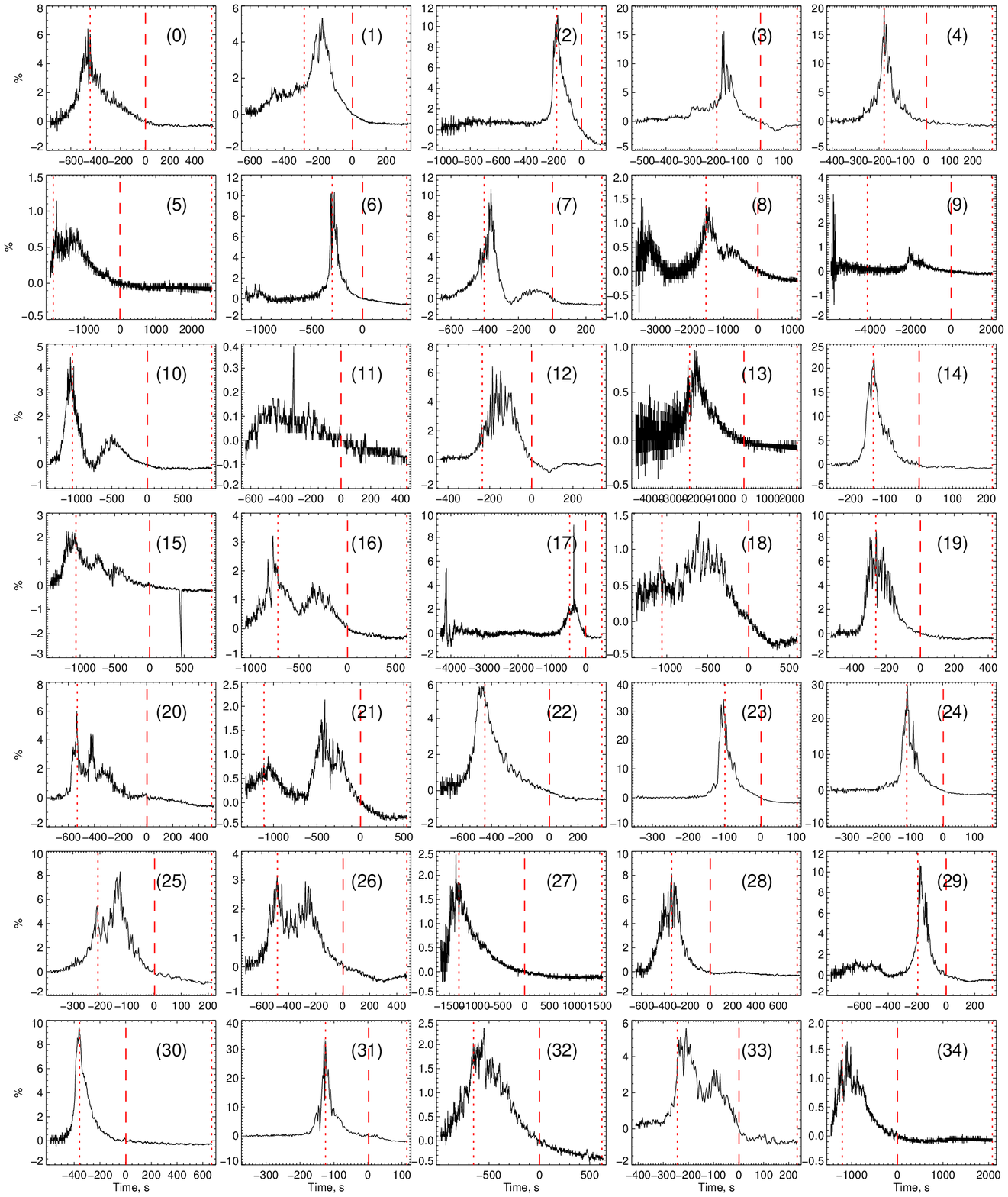}}
\caption{{ Time derivative of} GOES 1--8~\AA~ time-series normalised to the irradiance values{, showing the wide variety of patterns but the general presence of the QPP in the impulsive phase}.
Vertical dashed lines indicate the 1--8~\AA~peak and dotted lines indicate the arbitrarily defined flare onset (5\% of the maximum flux density) and end times.}
\label{fig:confuso1}
\end{figure}

We show { time derivative of the} timeseries plots for each of the 35 events in Figure~\ref{fig:confuso1}, normalised to the irradiance values. From the wavelet analysis, we found an average {characteristic timescale} of 16--53 seconds (at 95\% confidence level) { and 22--42 seconds (at 99.7\%)} (Table \ref{tab:bigtable}). { Overall, periodic signals in the range 8--112 seconds (at a confidence level of 99.7\%) appear across our flare sample.} For most of the events, we detected similar characteristic timescales in both channels during the rise {impulsive} phase. 
During the gradual phase, only 7 flares (20\% of the sample) had QPPs detected in the average wavelet power spectrum, although {some events gave the appearance of weak broad-band power in the accessible time range}. 
Moreover, during the impulsive phase of some events (indicated in the Table \ref{tab:bigtable} by $\dagger$), the average WPS level of the low channel was below the red-noise 95\% confidence level, yet there are significant regions clustered together in time and period, indicating processes with less random nature. For those cases, the high channel average WPS confirmed the presence of quasi-periodic power { at the confidence level of 95\%; however for these 7 events, at 99.7\% not much persistent quasi-periodic power was found.}

\begin{table}[htbp]
\caption{X-class flares of Cycle 24, along with the results of the wavelet analysis, { where we show the characteristic time scale of the WPS above the red-noise level, for confidence levels of 95\% and 99.7\%.} $\dagger$ indicate events with quasi-periodic power found locally but not evident in the average WPS.}\label{tab:bigtable}
\begin{tabular}{|r|c|r|r|r|c|}
\hline 
No. & Flare ID & Class &  \multicolumn{2}{|r|}{Characteristic  timescale}  & QPP\\
      &            &          & 95\% & 99.7\% & gradual \\
\hline
       0 & SOL2011-02-15 & X2.2 &  10-30  & 12-24 & no\\
       1 & SOL2011-03-09 & X1.5 &  10-30$^\dagger$ &  - & no\\
       2 & SOL2011-08-09 & X6.9 &  10-55  & 14-46&  yes\\
       3 & SOL2011-09-06 & X2.1 &  10-25  & 12-23&  no\\
       4 & SOL2011-09-07 & X1.8 &  10-45  & 10-37&  no\\
       5 & SOL2011-09-22 & X1.4 &  25-80$^\dagger$   & -& no\\
       6 & SOL2011-09-24 & X1.9 &  10-40  & 10-34&  no\\
       7 & SOL2011-11-03 & X1.9 &  10-25  & -& no \\
       8 & SOL2012-01-27 & X1.7 &  50-200  & 65-99&  yes\\
       9 & SOL2012-03-05 & X1.1 &  25-70  & 30-54&  yes\\
      10 & SOL2012-03-07 & X5.4 &  15-50  & 20-26&  no\\
      11 & SOL2012-03-07 & X1.3 &  30-55$^\dagger$   & -& no\\
      12 & SOL2012-07-06 & X1.1 &  10-30  & 12-25& no\\
      13 & SOL2012-07-12 & X1.4 &  9-20,50-200  & 70-106& yes\\
      14 & SOL2012-10-23 & X1.8 &  10-20$^\dagger$   & 10-14& no\\
      15 & SOL2013-05-13 & X1.7 &  10-60  & 17-45& no \\
      16 & SOL2013-05-13 & X2.8 &  15-40   & 19-31& yes\\
      17 & SOL2013-05-14 & X3.2 &  10-90  &16-29&  yes\\
      18 & SOL2013-05-15 & X1.2 &  30-70  & 35-57&  no\\
      19 & SOL2013-10-25 & X1.7 &  10-45  & 11-25&  no\\
      20 & SOL2013-10-25 & X2.1 &  10-25  & 10-18&  no\\
      21 & SOL2013-10-28 & X1.0 &  15-65  &17-24& no\\
      22 & SOL2013-10-29 & X2.3 &  20-65  & 24-54&  no\\
      23 & SOL2013-11-05 & X3.3 &  9-12$^\dagger$  &-&  no\\
      24 & SOL2013-11-08 & X1.1 &  8-20 &10-18&  no\\
      25 & SOL2013-11-10 & X1.1 &  9-25  &18-22&  no\\
      26 & SOL2013-11-19 & X1.0 &  10-30  & 12-26&  no\\
      27 & SOL2014-01-07 & X1.2 &  9-18$^\dagger$  & -&  no\\
      28 & SOL2014-02-25 & X4.9 &  8-45   & 12-40& no\\
      29 & SOL2014-03-29 & X1.0 &  8-60  & 8-56&  no\\
      30 & SOL2014-04-25 & X1.3 &  15-54  & 17-48&  yes\\
      31 & SOL2014-06-10 & X2.2 &  24-45$^\dagger$  & -&  no\\
      32 & SOL2014-06-10 & X1.5 &  22-82  & 26-35& no\\
      33 & SOL2014-06-11 & X1.0 &  10-16 &11-14&  no\\
      34 & SOL2014-09-10 & X1.6 &  28-126  & 80-112&  no\\
\hline
\end{tabular}
\end{table}

\subsection{Association with hard X-rays}

\begin{figure}[htbp]
{\includegraphics[width=0.49\textwidth]{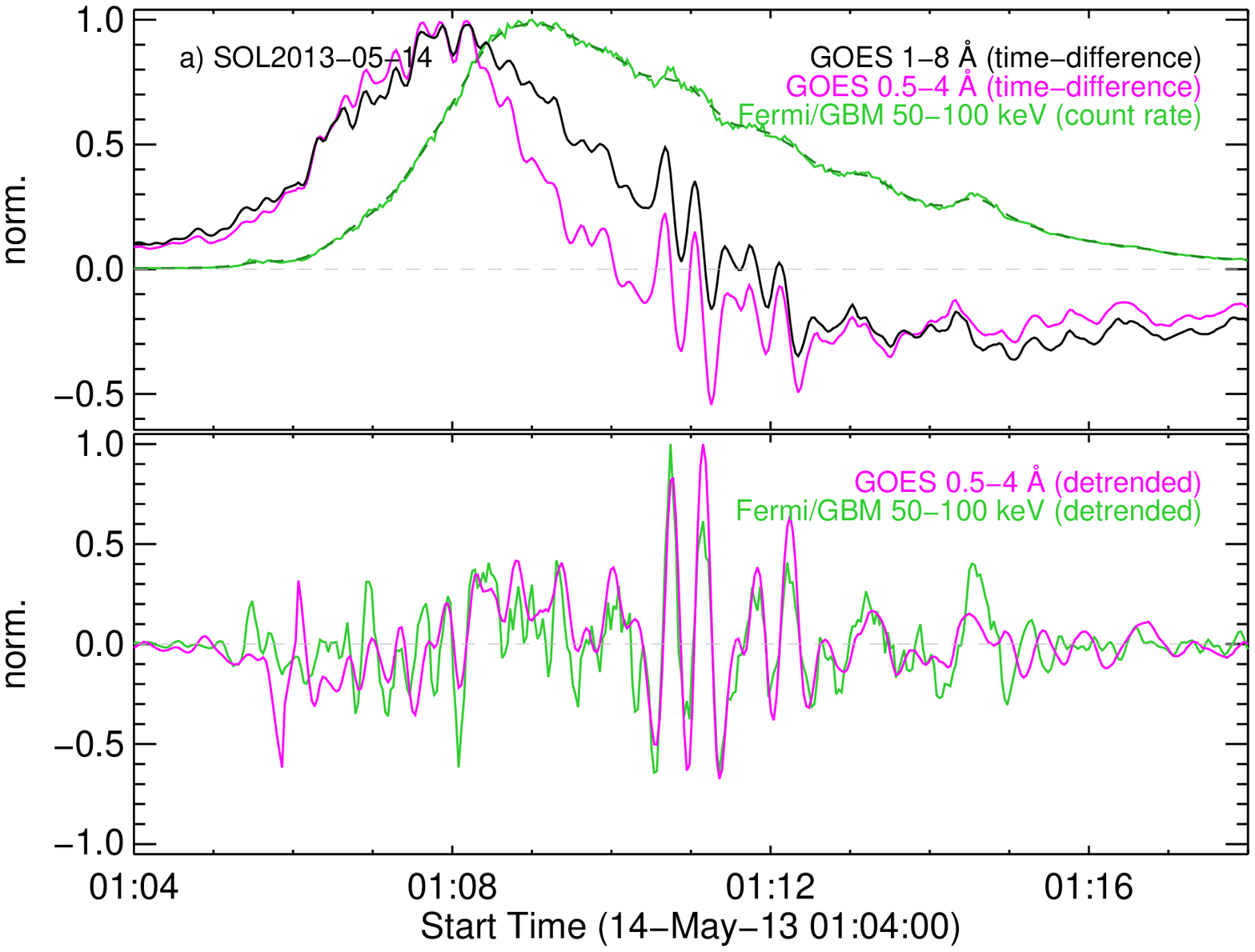}}
{\includegraphics[width=0.49\textwidth]{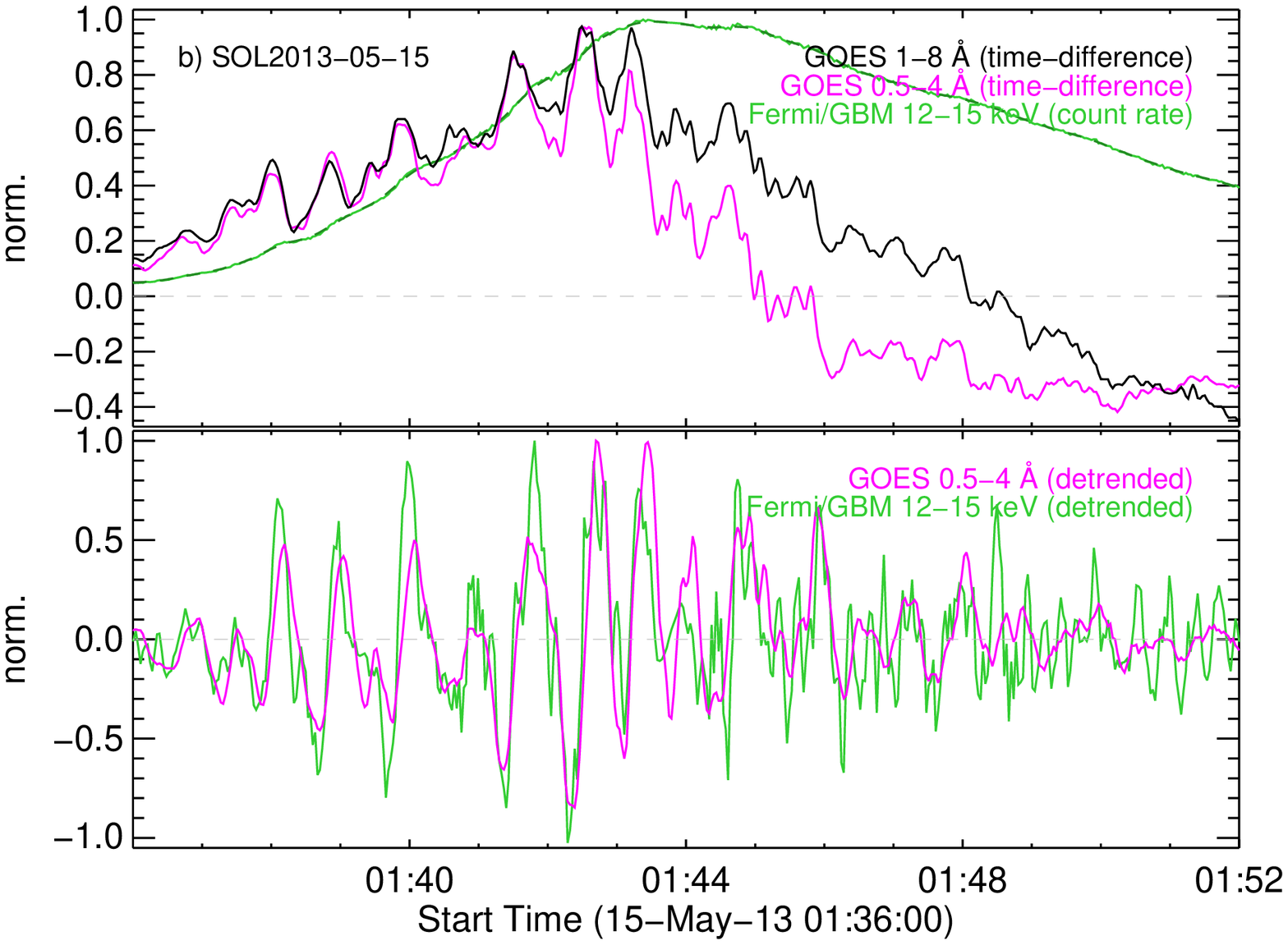}}
{\includegraphics[width=0.49\textwidth]{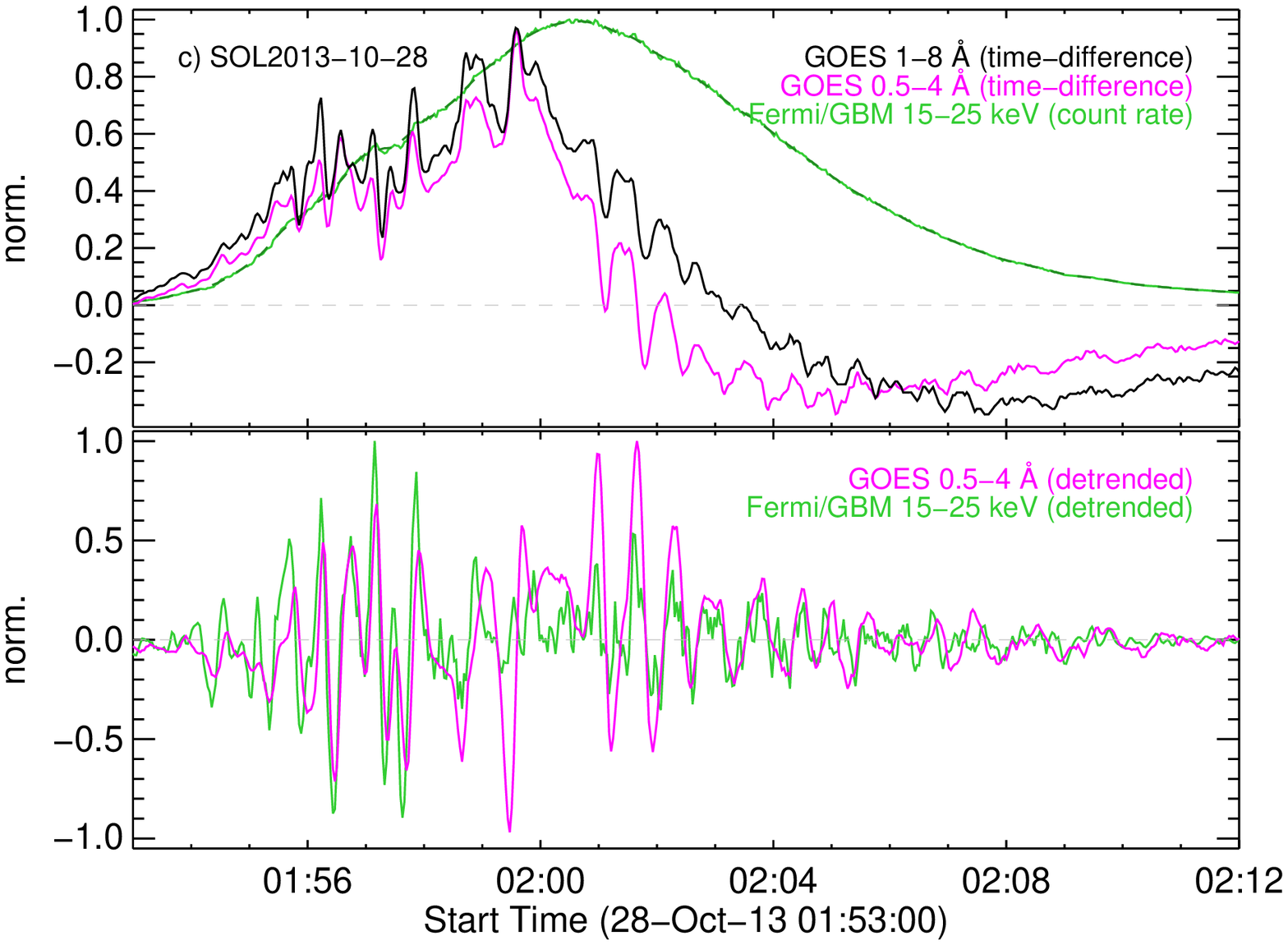}}
{\includegraphics[width=0.49\textwidth]{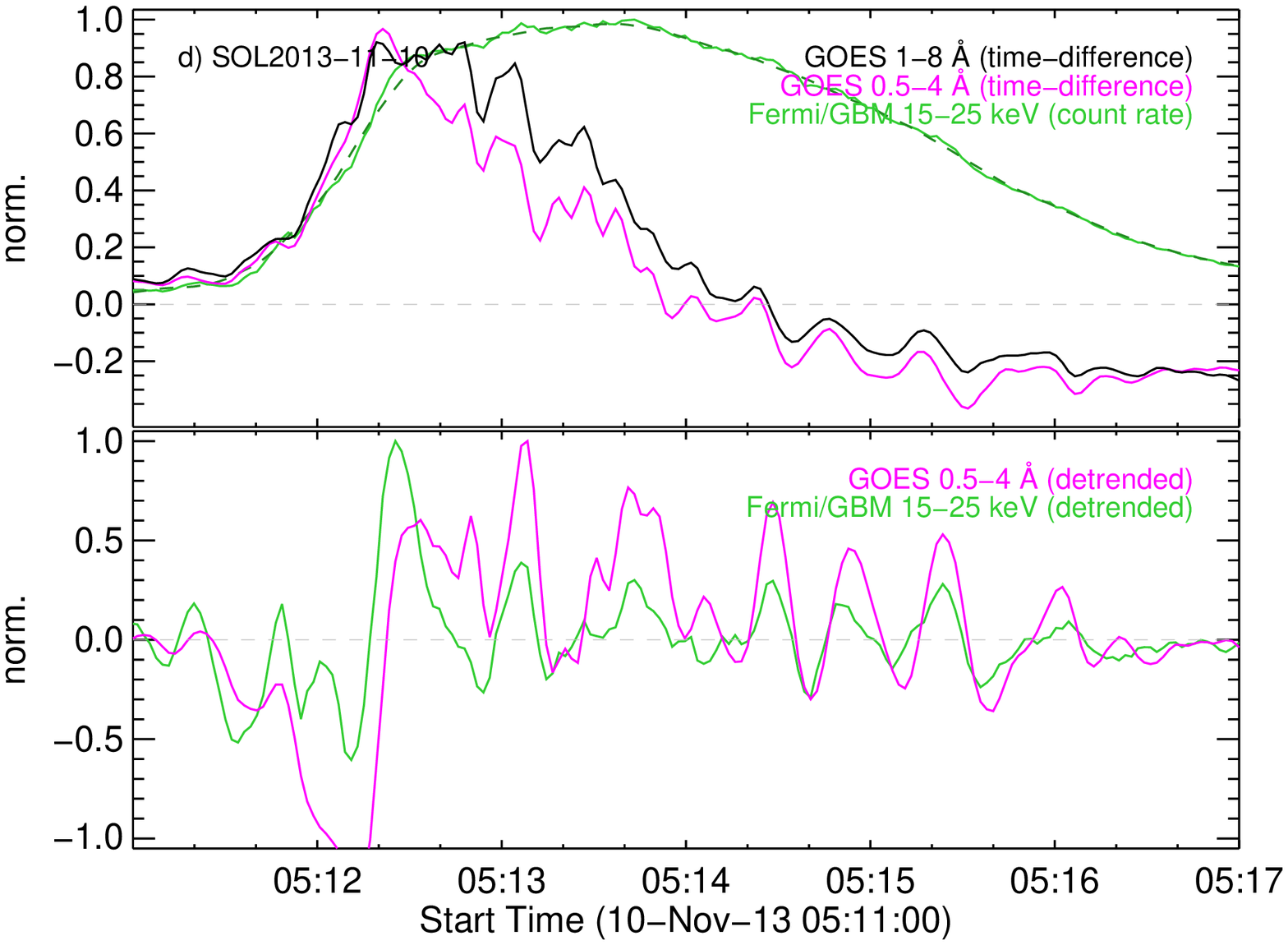}}
{\includegraphics[width=0.49\textwidth]{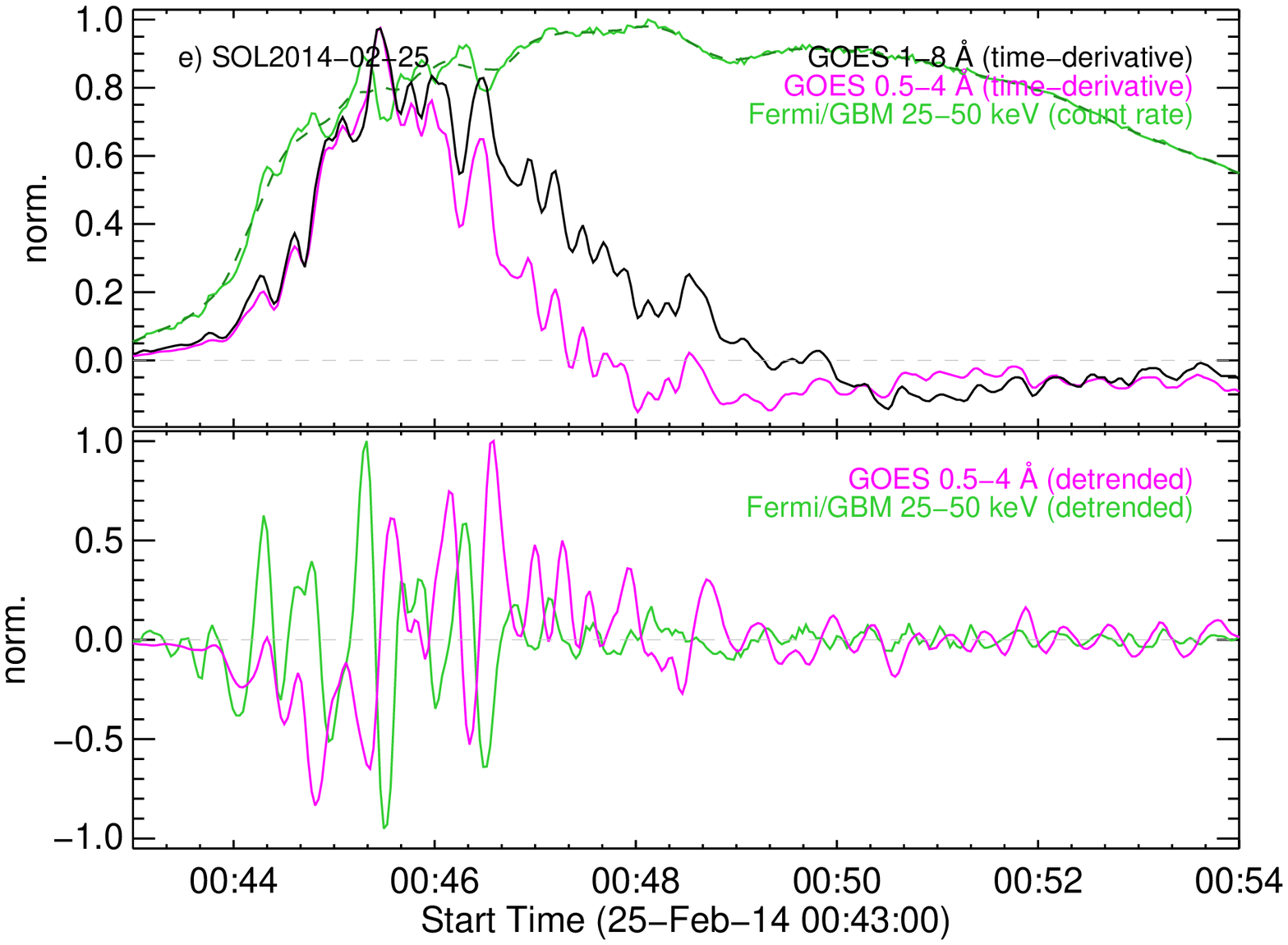}}
{\includegraphics[width=0.49\textwidth]{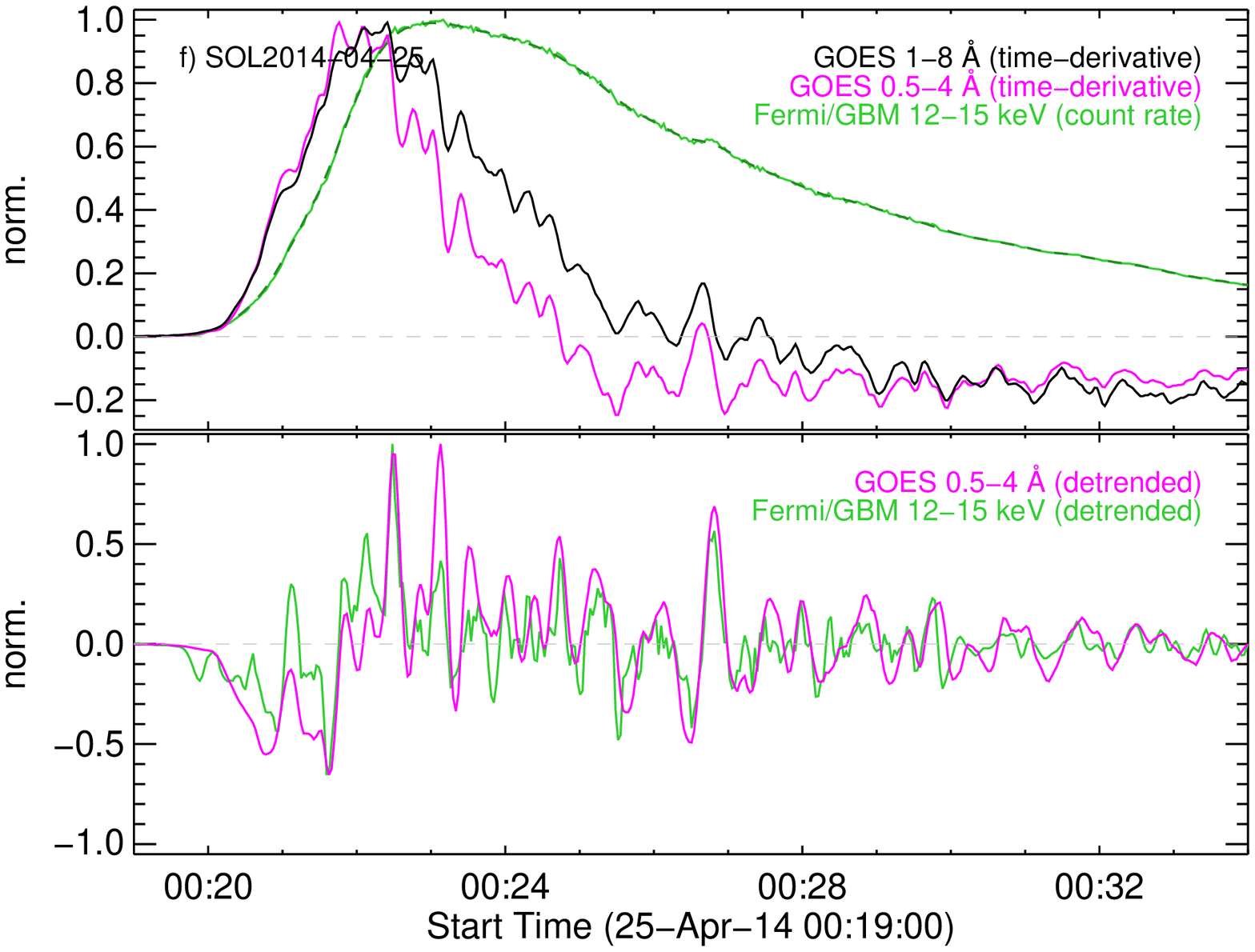}}
\caption{SXR (GOES) and HXR (\textit{Fermi}/GBM) pulsations for a subset of our sample. Each panel shows the { time derivative of} time-series of both GOES channels (for visualisation, as the detrended signals were used for comparison with the HXRs) and the \textit{Fermi}/GBM energy band with the highest cross-correlation coefficient (see Table \ref{tab:smalltable}), and the detrended time-series of \textit{Fermi}/GBM and GOES 0.5--4 \AA.
The upper right panel (SOL2013-05-15) shows an excellent pulse-by-pulse match between GOES and the hard X-rays, limited by the decreasing SNR at late times.
The only poor example is SOL2014-02-25 (lower left) in this set.}
\label{fig:hxr}
\end{figure}

For a subset of six of  our sample, we compared the SXR timeseries with hard X-rays (HXR) data from \textit{Fermi}/GBM 
\citep{Meegan:2009,2011AIPC.1366..155P,2010AAS...21640406S}. 
We detrended the intensity signals by subtracting their smoothed component using a boxcar of 30~s (note that \cite{2012ApJ...749L..16D} used 20~s). 
The { time derivative} of SXR and HXR lightcurves, and detrended components are shown in the Figure \ref{fig:hxr}. 
The SXR and HXR detrended signals were cross-correlated and the results found are in the Table \ref{tab:smalltable}. 
We found generally positive correlations between the SXR and HXR pulsations, with cross-correlation coefficient $r$ frequently greater than 0.5. 
For these cases of $r>0.5$ no delays were found between the signals. 
For events 18, 21, 25 and 30 (see Table \ref{tab:smalltable}) we found good correlation for many HXR energy bands; for the event 17, good correlation appeared for the higher energy channels ($>25$ keV), and finally, for event 28, we get $r \approx 0.4$ at all energy channels, with delays 12--20 sec. 
{This particular example, SOL2014-02-25, actually exhibited periods with anticorrelation between { detrended} $>25$~keV and the GOES.
This may result from a well-known artifact present in many scintillation-counter spectrometers operating at excessively high count rates, as explained below.
We note that this particular flare was an unusual and powerful $\gamma$-ray event that exhibited the sustained emission of $>100$~MeV photons
\citep[\textit{e.g.}][]{2014ApJ...787...15A}.
The correlations, though not perfect, point to a typical situation in which the QPP excitations detected in the GOES time-series closely match those seen in hard X-rays.}

{ We note that applying a filter (\textit{e.g.} Fourier) to remove the slow-varying component would have the same issues as the detrending technique, \textit{i.e.} the possibility of introducing spurious signals. We chose the latter as 1) it is easily reproducible and 2) it is the preferred method of the majority of authors investigating QPPs, and in particular, \cite{2012ApJ...749L..16D}, making it easy to compare the results of different authors. Moreover, our point here is to show the correlation of the variability at small amplitudes of soft and hard X-ray data and the implications of this, and not to perform a strict periodicity analysis. Periodic or not, such variability is real (as noted by \cite[\textit{e.g.}][]{2011A&A...533A..61G,InglisIrelandDominique:2015} and possibly affecting a wide range of the flare emission spectrum, as shown here and by \cite{2012ApJ...749L..16D}.}

\begin{table}[htbp]
\caption{Cross-correlation results between GOES SXR 0.5--4~\AA~ and \textit{Fermi}/GBM HXR channels (in keV).}\label{tab:smalltable}
\begin{tabular}{l|l|c|r|r|r|r|r|r}
\hline 
No. & Flare ID & NaI det. & 4--9 & 9--12 & 12--15 & 15--25 & 25--50 & 50--100 \\
\hline
17 & SOL2013-05-14 &n1& 0.25 & 0.19 & 0.23 & 0.22 & 0.42 & 0.75 \\ 

18 & SOL2013-05-15 &n4& 0.30 & 0.73 & 0.75 & 0.71 & 0.65 & 0.54 \\

21 & SOL2013-10-28 &n1& 0.67 & 0.29 & 0.31 & 0.67 & 0.61 & 0.46 \\

25 & SOL2013-11-10 &n4& 0.59 & 0.62 & 0.59 & 0.67 & 0.65 & 0.33 \\

  28 & SOL2014-02-25 &n4& 0.34 & 0.38 & 0.37 & 0.40 & 0.44 & 0.40 \\ 

30 & SOL2014-04-25 &n4& 0.38 & 0.72 & 0.75 & 0.67 & 0.37 & 0.22 \\ 
\hline
\end{tabular}
\end{table}

{As an important caveat regarding the hard X-ray data from \textit{Fermi}, we note that in many cases the archived lightcurves clearly exhibit nonlinear responses as a result of high counting rates (note that our sample contains specifically just the most energetic events). 
We have not tried to understand these effects in detail and are unaware of any published analysis of such effects in the \textit{Fermi} data, but past experience with scintillation counters at high rates suggests that both the spectrum and the timeseries can suffer.
For example, at high rates the photomultiplier gain may change, affecting the energy calibration of the detectors. { We have not used the most sunward NaI detectors to avoid these effects (see Table \ref{tab:smalltable} for detector used in this analysis); however, taking data from detectors with larger offset angles reduce the count rates at high energies and also affects the detection of the small scale pulsations.}
We note however, that during some parts of SOL2014-02-25's  time history, the impulsive energy signatures appear to change phase by 180$^\circ$, becoming minima (at some energies) rather than maxima; this would have a natural explanation in terms of gain changes. { Nevertheless, for this event we found similar results for different \textit{Fermi}/GBM NaI detectors, suggesting that the lack of correlation between the SXR and HXR may not be an instrumental effect.}}

\section{Discussion}

We have found that most of the X-class flares in our sample show clear pulsation signatures in their time-series, as would be expected from the presence of spikes in hard X-ray emission \citep[\textit{e.g.}][]{1993SoPh..146..177D} and the operation of the Neupert effect on these time scales. 
These signatures appear strongly in the impulsive phase of the flare, but much less prominently in the gradual phase. 

Our analysis typically covers the frequency range about four octaves below the Nyquist frequency of the sampling, { in the band} 0.01--0.25~Hz. Although the GOES peaks may have time scales down to our limit of 4~s, most of the variability is slower. 
The wavelet analysis for the most part does not show narrow features in spite of the presence of { multiple peaks in the time-series, and therefore need not be interpreted in terms of a resonant process}.
This seems reasonable, given the rapid structural changes of the emitting regions during the impulsive phase; one would not expect stable eigenfrequencies to dominate the time signatures.
Later on and remote from the site of energy release, distinct signatures of damped oscillations appear commonly \citep{1999ApJ...520..880A}.

{ The QPP variability found here extends into the 0.1~Hz band, as discovered by \cite{1969ApJ...155L.117P}, but involves longer time scales (8-112 seconds) and are in overall agreement with the periods found in many QPP studies in different spectral ranges in the literature \citep[\textit{e.g.}][]{2003A&A...412L...7N,2005A&A...439..727M,2007ARep...51..588R,2008A&A...487.1147I,InglisNakariakov:2009,2012SoPh..280..491K}. Parks and Winckler identified a specific time scale (16~$s$, or roughly 60~mHz), and we find a few cases in which there is a spectral peak in the range above about 80~mHz (or about 12 seconds). Curiously one of the best-studied recent QPP events, SOL1998-05-08, shows almost exactly the same time scale \citep{2001ApJ...562L.103A,2004AstL...30..480S,2008A&A...487.1147I,InglisIrelandDominique:2015}, and is one of the best examples of QPP that could arise from a true oscillation of some sort. \cite{Kupriyanova:2010} investigated QPPs at 17 GHz using data from the \textit{Nobeyama Radio Heliograph} and found, for ten events, periods in the range 5--60 seconds. In particular, in our analysis of SOL2011-02-15 (see Figure \ref{fig:wavelet}) we found the characteristic time scale of 12--24 seconds, in agreement with the results of \cite{2012ApJ...749L..16D}.We note that \cite{InglisIrelandDominique:2015} periodicity analysis of this event, using several HXR energy bands from \textit{Fermi}/GBM and extreme ultraviolet (EUV) data from PROBA2/LYRA, did not find a periodic signal. We argue that analyzing the global wavelet power or Fourier power \citep[\textit{e.g.}][]{InglisIrelandDominique:2015} seems to reduce the chances to detect any periodic power because the signal is averaged over the entire duration of the events. Any periodic signal would have to be persistently strong through the entire flare in order to appear. Moreover, the variability in the GOES SXR time series has a very small amplitude in comparison to the slow-varying component thus a filter must be applied to retrieve it.}

\begin{figure}[htbp]
\centering{\includegraphics[width=0.65\textwidth]{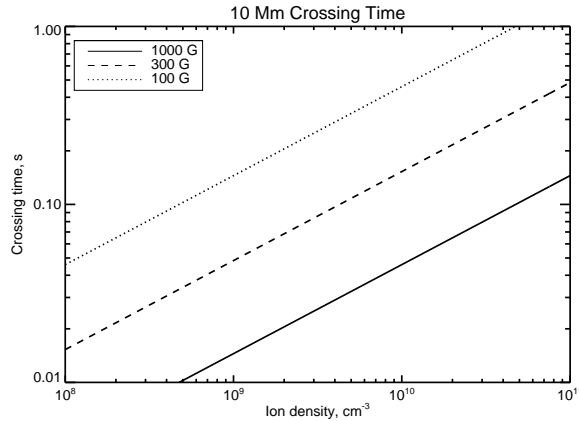}}
\caption{Estimates of the Alfv{\' e}n time scale for {coronal/chromospheric structures for ion densities in the range $10^{8-11}$~cm$^{-3}$ and for three plausible values of the magnetic field intensity, for a 10~Mm scale.}
}
\label{fig:alfven_time}
\end{figure}

The observed time scales are therefore inconsistent with Alfv{\' e}nic processes in the core of an active region, as reflected in Figure~\ref{fig:alfven_time} \citep[see also][]{2008ApJ...675.1645F}. In the impulsive phase of a flare, we expect low plasma densities since evaporation is just beginning to happen, and high magnetic fields since the events typically occur in active regions with sunspots. Thus the preferred domain in Figure~\ref{fig:alfven_time} would be the lower-right quadrant.
This suggests that the { chromospheric/photospheric regions} of the flares may help to define the time scales of the flare development, or the travel times of the QPP disturbances, since the deep atmosphere has slower characteristic time scales.

The soft X-ray variability is maximal at the very beginning (Figure \ref{fig:confuso1}), which we readily explain by 
the accumulation of hot material in the corona, which systematically tends to obscure new contributions.

\section{Conclusions}

We have surveyed the X-class flares of the current solar cycle (35 events from SOL2011-02-15 to SOL2014-09-10), as viewed by the standard soft X-ray irradiance measurements from the GOES satellites.
Over this time interval the GOES-15 detectors provided the data, and (as described in Section~\ref{sec:goes}) its sampling precision greatly exceeds that of earlier members of the GOES series.
This and the excellent signal-to-noise ratio of the observations allowed us to use { digital time derivative of the} time-series to study the variability in detail, up to the Nyquist frequency of the sampling at 0.25~Hz.
The main conclusion of this paper is, therefore, to confirm the \cite{2012ApJ...749L..16D} result that the recent GOES measurements work very well in studies of QPP.
By ``recent'' we mean { GOES 13-15}, with coverage from September 2010, and therefore the source of all of the data on X-class flares in Cycle~24 thus far.
These data have better precision than that of earlier GOES instruments, but still do not adequately sample the detector noise (Section~\ref{sec:goes}).

Using the { time derivative of} time-series, we find that { 28 out of 35} X-class flares show QPP to some degree, especially during the impulsive phase of the event development. { Here we define the presence of QPP as the identification of power in the wavelet power spectrum (averaged over the impulsive phase) above the red-noise model, with a confidence level of 99.7\%}. Overall, the characteristic time scale found is in the range 8--112 seconds (see Table \ref{tab:bigtable}).
Because of the sampling issue we have discussed, the data are unable clearly to show QPP during the gradual phases of the flares.
We find that most of the spectral power typically is at lower frequencies { (between roughly 8--112 seconds)} and is explicitly inconsistent with signal propagation at reasonable values of the Alfv{\' e}n times for relevant scales in the active-region corona and upper chromosphere.

The QPP phenomenon in solar flares could be interpreted in several ways \citep[\textit{e.g.}][]{2009SSRv..149..119N}.
\cite{2011ApJ...730L..27N}, for example, describe a scenario in which slow-mode MHD waves trigger episodes of magnetic reconnection at successive locations.
We think this unlikely for the impulsive phase of a flare; the presence of non-steady evolution of the magnetic field, as required to release magnetic energy rapidly, would tend to destroy the identities of eigenmodes as the impulsive phase develops.
Such evolution is not typically considered in theoretical work in the area of coronal seismology,  which is largely based on a the structure of a fixed cylinder in the absence of non-steady flows \citep{2005LRSP....2....3N}.
The problem of time scales also remains: Nakariakov and Zimovets hypothesize slow coupling between arcade loops via perpendicular transport in the footpoint regions, but there would be no obvious source of resonant behavior or periodicity in such a process. 

%

%

%
 \begin{acks}
We would like to thank the anonymous reviwer for the comments and suggestions that helped to improve the paper. We would like to thank Marie Dominique and Rodney Viereck for helpful discussions.
Author HSH thanks NASA for support under contract NAS 5-98033 for RHESSI, and acknowledges warm hospitality at the University of Glasgow. The research leading to these results has received funding from the European Community's Seventh Framework Programme (FP7/2007-2013) under grant agreement no. 606862 (F-CHROMA) (PJAS, LF), from STFC grant ST/I001808/1 (PJAS, LF) and ST/L000741/1 (LF).
 \end{acks}

%
\newpage
\bibliographystyle{spr-mp-sola}
\bibliography{goes}  

\begin{thebibliography}{42}
\ifx\bisbn     \undefined \def\bisbn  #1{ISBN #1}\fi
\ifx\binits    \undefined \def\binits#1{#1}\fi
\ifx\bauthor   \undefined \def\bauthor#1{#1}\fi
\ifx\batitle   \undefined \def\batitle#1{#1}\fi
\ifx\bjtitle   \undefined \def\bjtitle#1{\textit{#1}}\fi
\ifx\bvolume   \undefined \def\bvolume#1{\textbf{#1}}\fi
\ifx\byear     \undefined \def\byear#1{#1}\fi
\ifx\bissue    \undefined \def\bissue#1{#1}\fi
\ifx\bfpage    \undefined \def\bfpage#1{#1}\fi
\ifx\blpage    \undefined \def\blpage #1{#1}\fi
\ifx\burl      \undefined \def\burl#1{\textsf{#1}}\fi
\ifx\href      \undefined \def\href#1#2{\textsf{#2}}\fi
\ifx\betal     \undefined \def\betal{\textit{et al.}}\fi
\ifx\bctitle   \undefined \def\bctitle#1{#1}\fi
\ifx\beditor   \undefined \def\beditor#1{#1}\fi
\ifx\bbtitle   \undefined \def\bbtitle#1{\textit{#1}}\fi
\ifx\bedition  \undefined \def\bedition#1{#1}\fi
\ifx\bseriesno \undefined \def\bseriesno#1{\textbf{#1}}\fi
\ifx\blocation \undefined \def\blocation#1{#1}\fi
\ifx\bsertitle \undefined \def\bsertitle#1{\textit{#1}}\fi
\ifx\bsnm      \undefined \def\bsnm#1{#1}\fi
\ifx\bsuffix   \undefined \def\bsuffix#1{#1}\fi
\ifx\bparticle \undefined \def\bparticle#1{#1}\fi
\ifx\barticle  \undefined \def\barticle#1{}\fi
\ifx\binstitute  \undefined \def\binstitute#1{#1}\fi
\ifx\bpublisher  \undefined \def\bpublisher#1{#1}\fi
\ifx\doiurl    \undefined
  \def\doiurl#1{\href{http://dx.doi.org/#1}{\textsf{DOI}}}\fi
\ifx\arxivurl  \undefined
  \def\arxivurl#1{\href{http://arxiv.org/abs/#1}{\textsf{arXiv}}}\fi
\ifx\adsurl    \undefined
  \def\adsurl#1{\href{http://adsabs.harvard.edu/abs/#1}{\textsf{ADS}}}\fi
\ifx\botherref \undefined \def\botherref#1{}\fi
\ifx\url       \undefined \def\url#1{\textsf{#1}}\fi
\ifx\bchapter  \undefined \def\bchapter#1{}\fi
\ifx\bbook     \undefined \def\bbook#1{}\fi
\ifx\bcomment  \undefined \def\bcomment#1{#1}\fi
\ifx\oauthor   \undefined \def\oauthor#1{#1}\fi
\ifx\citeauthoryear \undefined\def \citeauthoryear#1{#1}\fi
\def\endbibitem {}
\ifx\bconflocation  \undefined \def\bconflocation#1{#1} \fi

\bibitem[\protect\citeauthoryear{{Ackermann}
  \textit{et~al.}}{2014}]{2014ApJ...787...15A}
\begin{barticle}
\bauthor{\bsnm{{Ackermann}}, \binits{M.}},
\bauthor{\bsnm{{Ajello}}, \binits{M.}},
\bauthor{\bsnm{{Albert}}, \binits{A.}},
\bauthor{\bsnm{{Allafort}}, \binits{A.}},
\bauthor{\bsnm{{Baldini}}, \binits{L.}},
\bauthor{\bsnm{{Barbiellini}}, \binits{G.}},
\bauthor{\bsnm{{Bastieri}}, \binits{D.}},
\bauthor{\bsnm{{Bechtol}}, \binits{K.}},
\bauthor{\bsnm{{Bellazzini}}, \binits{R.}},
\bauthor{\bsnm{{Bissaldi}}, \binits{E.}},
\bauthor{\bsnm{{Bonamente}}, \binits{E.}},
\bauthor{\bsnm{{Bottacini}}, \binits{E.}},
\bauthor{\bsnm{{Bouvier}}, \binits{A.}},
\bauthor{\bsnm{{Brandt}}, \binits{T.J.}},
\bauthor{\bsnm{{Bregeon}}, \binits{J.}},
\bauthor{\bsnm{{Brigida}}, \binits{M.}},
\bauthor{\bsnm{{Bruel}}, \binits{P.}},
\bauthor{\bsnm{{Buehler}}, \binits{R.}},
\bauthor{\bsnm{{Buson}}, \binits{S.}},
\bauthor{\bsnm{{Caliandro}}, \binits{G.A.}},
\bauthor{\bsnm{{Cameron}}, \binits{R.A.}},
\bauthor{\bsnm{{Caraveo}}, \binits{P.A.}},
\bauthor{\bsnm{{Cecchi}}, \binits{C.}},
\bauthor{\bsnm{{Charles}}, \binits{E.}},
\bauthor{\bsnm{{Chekhtman}}, \binits{A.}},
\bauthor{\bsnm{{Chen}}, \binits{Q.}},
\bauthor{\bsnm{{Chiang}}, \binits{J.}},
\bauthor{\bsnm{{Chiaro}}, \binits{G.}},
\bauthor{\bsnm{{Ciprini}}, \binits{S.}},
\bauthor{\bsnm{{Claus}}, \binits{R.}},
\bauthor{\bsnm{{Cohen-Tanugi}}, \binits{J.}},
\bauthor{\bsnm{{Conrad}}, \binits{J.}},
\bauthor{\bsnm{{Cutini}}, \binits{S.}},
\bauthor{\bsnm{{D'Ammando}}, \binits{F.}},
\bauthor{\bsnm{{de Angelis}}, \binits{A.}},
\bauthor{\bsnm{{de Palma}}, \binits{F.}},
\bauthor{\bsnm{{Dermer}}, \binits{C.D.}},
\bauthor{\bsnm{{Desiante}}, \binits{R.}},
\bauthor{\bsnm{{Digel}}, \binits{S.W.}},
\bauthor{\bsnm{{Di Venere}}, \binits{L.}},
\bauthor{\bsnm{{Silva}}, \binits{E.d.C.e.}},
\bauthor{\bsnm{{Drell}}, \binits{P.S.}},
\bauthor{\bsnm{{Drlica-Wagner}}, \binits{A.}},
\bauthor{\bsnm{{Favuzzi}}, \binits{C.}},
\bauthor{\bsnm{{Fegan}}, \binits{S.J.}},
\bauthor{\bsnm{{Focke}}, \binits{W.B.}},
\bauthor{\bsnm{{Franckowiak}}, \binits{A.}},
\bauthor{\bsnm{{Fukazawa}}, \binits{Y.}},
\bauthor{\bsnm{{Funk}}, \binits{S.}},
\bauthor{\bsnm{{Fusco}}, \binits{P.}},
\bauthor{\bsnm{{Gargano}}, \binits{F.}},
\bauthor{\bsnm{{Gasparrini}}, \binits{D.}},
\bauthor{\bsnm{{Germani}}, \binits{S.}},
\bauthor{\bsnm{{Giglietto}}, \binits{N.}},
\bauthor{\bsnm{{Giordano}}, \binits{F.}},
\bauthor{\bsnm{{Giroletti}}, \binits{M.}},
\bauthor{\bsnm{{Glanzman}}, \binits{T.}},
\bauthor{\bsnm{{Godfrey}}, \binits{G.}},
\bauthor{\bsnm{{Grenier}}, \binits{I.A.}},
\bauthor{\bsnm{{Grove}}, \binits{J.E.}},
\bauthor{\bsnm{{Guiriec}}, \binits{S.}},
\bauthor{\bsnm{{Hadasch}}, \binits{D.}},
\bauthor{\bsnm{{Hayashida}}, \binits{M.}},
\bauthor{\bsnm{{Hays}}, \binits{E.}},
\bauthor{\bsnm{{Horan}}, \binits{D.}},
\bauthor{\bsnm{{Hughes}}, \binits{R.E.}},
\bauthor{\bsnm{{Inoue}}, \binits{Y.}},
\bauthor{\bsnm{{Jackson}}, \binits{M.S.}},
\bauthor{\bsnm{{Jogler}}, \binits{T.}},
\bauthor{\bsnm{{J{\'o}hannesson}}, \binits{G.}},
\bauthor{\bsnm{{Johnson}}, \binits{W.N.}},
\bauthor{\bsnm{{Kamae}}, \binits{T.}},
\bauthor{\bsnm{{Kawano}}, \binits{T.}},
\bauthor{\bsnm{{Kn{\"o}dlseder}}, \binits{J.}},
\bauthor{\bsnm{{Kuss}}, \binits{M.}},
\bauthor{\bsnm{{Lande}}, \binits{J.}},
\bauthor{\bsnm{{Larsson}}, \binits{S.}},
\bauthor{\bsnm{{Latronico}}, \binits{L.}},
\bauthor{\bsnm{{Lemoine-Goumard}}, \binits{M.}},
\bauthor{\bsnm{{Longo}}, \binits{F.}},
\bauthor{\bsnm{{Loparco}}, \binits{F.}},
\bauthor{\bsnm{{Lott}}, \binits{B.}},
\bauthor{\bsnm{{Lovellette}}, \binits{M.N.}},
\bauthor{\bsnm{{Lubrano}}, \binits{P.}},
\bauthor{\bsnm{{Mayer}}, \binits{M.}},
\bauthor{\bsnm{{Mazziotta}}, \binits{M.N.}},
\bauthor{\bsnm{{McEnery}}, \binits{J.E.}},
\bauthor{\bsnm{{Michelson}}, \binits{P.F.}},
\bauthor{\bsnm{{Mizuno}}, \binits{T.}},
\bauthor{\bsnm{{Moiseev}}, \binits{A.A.}},
\bauthor{\bsnm{{Monte}}, \binits{C.}},
\bauthor{\bsnm{{Monzani}}, \binits{M.E.}},
\bauthor{\bsnm{{Moretti}}, \binits{E.}},
\bauthor{\bsnm{{Morselli}}, \binits{A.}},
\bauthor{\bsnm{{Moskalenko}}, \binits{I.V.}},
\bauthor{\bsnm{{Murgia}}, \binits{S.}},
\bauthor{\bsnm{{Murphy}}, \binits{R.}},
\bauthor{\bsnm{{Nemmen}}, \binits{R.}},
\bauthor{\bsnm{{Nuss}}, \binits{E.}},
\bauthor{\bsnm{{Ohno}}, \binits{M.}},
\bauthor{\bsnm{{Ohsugi}}, \binits{T.}},
\bauthor{\bsnm{{Okumura}}, \binits{A.}},
\bauthor{\bsnm{{Omodei}}, \binits{N.}},
\bauthor{\bsnm{{Orienti}}, \binits{M.}},
\bauthor{\bsnm{{Orlando}}, \binits{E.}},
\bauthor{\bsnm{{Ormes}}, \binits{J.F.}},
\bauthor{\bsnm{{Paneque}}, \binits{D.}},
\bauthor{\bsnm{{Panetta}}, \binits{J.H.}},
\bauthor{\bsnm{{Perkins}}, \binits{J.S.}},
\bauthor{\bsnm{{Pesce-Rollins}}, \binits{M.}},
\bauthor{\bsnm{{Petrosian}}, \binits{V.}},
\bauthor{\bsnm{{Piron}}, \binits{F.}},
\bauthor{\bsnm{{Pivato}}, \binits{G.}},
\bauthor{\bsnm{{Porter}}, \binits{T.A.}},
\bauthor{\bsnm{{Rain{\`o}}}, \binits{S.}},
\bauthor{\bsnm{{Rando}}, \binits{R.}},
\bauthor{\bsnm{{Razzano}}, \binits{M.}},
\bauthor{\bsnm{{Reimer}}, \binits{A.}},
\bauthor{\bsnm{{Reimer}}, \binits{O.}},
\bauthor{\bsnm{{Ritz}}, \binits{S.}},
\bauthor{\bsnm{{Schulz}}, \binits{A.}},
\bauthor{\bsnm{{Sgr{\`o}}}, \binits{C.}},
\bauthor{\bsnm{{Siskind}}, \binits{E.J.}},
\bauthor{\bsnm{{Spandre}}, \binits{G.}},
\bauthor{\bsnm{{Spinelli}}, \binits{P.}},
\bauthor{\bsnm{{Takahashi}}, \binits{H.}},
\bauthor{\bsnm{{Takeuchi}}, \binits{Y.}},
\bauthor{\bsnm{{Tanaka}}, \binits{Y.}},
\bauthor{\bsnm{{Thayer}}, \binits{J.G.}},
\bauthor{\bsnm{{Thayer}}, \binits{J.B.}},
\bauthor{\bsnm{{Thompson}}, \binits{D.J.}},
\bauthor{\bsnm{{Tibaldo}}, \binits{L.}},
\bauthor{\bsnm{{Tinivella}}, \binits{M.}},
\bauthor{\bsnm{{Tosti}}, \binits{G.}},
\bauthor{\bsnm{{Troja}}, \binits{E.}},
\bauthor{\bsnm{{Tronconi}}, \binits{V.}},
\bauthor{\bsnm{{Usher}}, \binits{T.L.}},
\bauthor{\bsnm{{Vandenbroucke}}, \binits{J.}},
\bauthor{\bsnm{{Vasileiou}}, \binits{V.}},
\bauthor{\bsnm{{Vianello}}, \binits{G.}},
\bauthor{\bsnm{{Vitale}}, \binits{V.}},
\bauthor{\bsnm{{Werner}}, \binits{M.}},
\bauthor{\bsnm{{Winer}}, \binits{B.L.}},
\bauthor{\bsnm{{Wood}}, \binits{D.L.}},
\bauthor{\bsnm{{Wood}}, \binits{K.S.}},
\bauthor{\bsnm{{Wood}}, \binits{M.}},
\bauthor{\bsnm{{Yang}}, \binits{Z.}}:
\byear{2014},
\batitle{{High-energy Gamma-Ray Emission from Solar Flares: Summary of Fermi
  Large Area Telescope Detections and Analysis of Two M-class Flares}}.
\bjtitle{\apj}
\bvolume{787},
\bfpage{15}.
\doiurl{10.1088/0004-637X/787/1/15}.
\adsurl{2014ApJ...787...15A}.
\end{barticle}
\endbibitem

\bibitem[\protect\citeauthoryear{{Asai}
  \textit{et~al.}}{2001}]{2001ApJ...562L.103A}
\begin{barticle}
\bauthor{\bsnm{{Asai}}, \binits{A.}},
\bauthor{\bsnm{{Shimojo}}, \binits{M.}},
\bauthor{\bsnm{{Isobe}}, \binits{H.}},
\bauthor{\bsnm{{Morimoto}}, \binits{T.}},
\bauthor{\bsnm{{Yokoyama}}, \binits{T.}},
\bauthor{\bsnm{{Shibasaki}}, \binits{K.}},
\bauthor{\bsnm{{Nakajima}}, \binits{H.}}:
\byear{2001},
\batitle{{Periodic Acceleration of Electrons in the 1998 November 10 Solar
  Flare}}.
\bjtitle{\apjl}
\bvolume{562},
\bfpage{L103}.
\doiurl{10.1086/338052}.
\adsurl{2001ApJ...562L.103A}.
\end{barticle}
\endbibitem

\bibitem[\protect\citeauthoryear{{Aschwanden}
  \textit{et~al.}}{1999}]{1999ApJ...520..880A}
\begin{barticle}
\bauthor{\bsnm{{Aschwanden}}, \binits{M.J.}},
\bauthor{\bsnm{{Fletcher}}, \binits{L.}},
\bauthor{\bsnm{{Schrijver}}, \binits{C.J.}},
\bauthor{\bsnm{{Alexander}}, \binits{D.}}:
\byear{1999},
\batitle{{Coronal Loop Oscillations Observed with the Transition Region and
  Coronal Explorer}}.
\bjtitle{\apj}
\bvolume{520},
\bfpage{880}.
\doiurl{10.1086/307502}.
\adsurl{1999ApJ...520..880A}.
\end{barticle}
\endbibitem

\bibitem[\protect\citeauthoryear{{Chamberlin}
  \textit{et~al.}}{2009}]{2009SPIE.7438E..02C}
\begin{bchapter}
\bauthor{\bsnm{{Chamberlin}}, \binits{P.C.}},
\bauthor{\bsnm{{Woods}}, \binits{T.N.}},
\bauthor{\bsnm{{Eparvier}}, \binits{F.G.}},
\bauthor{\bsnm{{Jones}}, \binits{A.R.}}:
\byear{2009},
\bctitle{{Next generation x-ray sensor (XRS) for the NOAA GOES-R satellite
  series}}.
In: \bbtitle{Society of Photo-Optical Instrumentation Engineers (SPIE)
  Conference Series},
\bsertitle{Society of Photo-Optical Instrumentation Engineers (SPIE) Conference
  Series}
\bseriesno{7438}.
\doiurl{10.1117/12.826807}.
\adsurl{2009SPIE.7438E..02C}.
\end{bchapter}
\endbibitem

\bibitem[\protect\citeauthoryear{{Dennis} and
  {Zarro}}{1993}]{1993SoPh..146..177D}
\begin{barticle}
\bauthor{\bsnm{{Dennis}}, \binits{B.R.}},
\bauthor{\bsnm{{Zarro}}, \binits{D.M.}}:
\byear{1993},
\batitle{{The Neupert effect - What can it tell us about the impulsive and
  gradual phases of solar flares?}}
\bjtitle{\solphys}
\bvolume{146},
\bfpage{177}.
\doiurl{10.1007/BF00662178}.
\adsurl{1993SoPh..146..177D}.
\end{barticle}
\endbibitem

\bibitem[\protect\citeauthoryear{{Dolla}
  \textit{et~al.}}{2012}]{2012ApJ...749L..16D}
\begin{barticle}
\bauthor{\bsnm{{Dolla}}, \binits{L.}},
\bauthor{\bsnm{{Marqu{\'e}}}, \binits{C.}},
\bauthor{\bsnm{{Seaton}}, \binits{D.B.}},
\bauthor{\bsnm{{Van Doorsselaere}}, \binits{T.}},
\bauthor{\bsnm{{Dominique}}, \binits{M.}},
\bauthor{\bsnm{{Berghmans}}, \binits{D.}},
\bauthor{\bsnm{{Cabanas}}, \binits{C.}},
\bauthor{\bsnm{{De Groof}}, \binits{A.}},
\bauthor{\bsnm{{Schmutz}}, \binits{W.}},
\bauthor{\bsnm{{Verdini}}, \binits{A.}},
\bauthor{\bsnm{{West}}, \binits{M.J.}},
\bauthor{\bsnm{{Zender}}, \binits{J.}},
\bauthor{\bsnm{{Zhukov}}, \binits{A.N.}}:
\byear{2012},
\batitle{{Time Delays in Quasi-periodic Pulsations Observed during the X2.2
  Solar Flare on 2011 February 15}}.
\bjtitle{\apjl}
\bvolume{749},
\bfpage{L16}.
\doiurl{10.1088/2041-8205/749/1/L16}.
\adsurl{2012ApJ...749L..16D}.
\end{barticle}
\endbibitem

\bibitem[\protect\citeauthoryear{{Fletcher} and
  {Hudson}}{2008}]{2008ApJ...675.1645F}
\begin{barticle}
\bauthor{\bsnm{{Fletcher}}, \binits{L.}},
\bauthor{\bsnm{{Hudson}}, \binits{H.S.}}:
\byear{2008},
\batitle{{Impulsive Phase Flare Energy Transport by Large-Scale Alfv{\'e}n W
  aves and the Electron Acceleration Problem}}.
\bjtitle{\apj}
\bvolume{675},
\bfpage{1645}.
\doiurl{10.1086/527044}.
\adsurl{2008ApJ...675.1645F}.
\end{barticle}
\endbibitem

\bibitem[\protect\citeauthoryear{{Fletcher}
  \textit{et~al.}}{2011}]{2011SSRv..159...19F}
\begin{barticle}
\bauthor{\bsnm{{Fletcher}}, \binits{L.}},
\bauthor{\bsnm{{Dennis}}, \binits{B.R.}},
\bauthor{\bsnm{{Hudson}}, \binits{H.S.}},
\bauthor{\bsnm{{Krucker}}, \binits{S.}},
\bauthor{\bsnm{{Phillips}}, \binits{K.}},
\bauthor{\bsnm{{Veronig}}, \binits{A.}},
\bauthor{\bsnm{{Battaglia}}, \binits{M.}},
\bauthor{\bsnm{{Bone}}, \binits{L.}},
\bauthor{\bsnm{{Caspi}}, \binits{A.}},
\bauthor{\bsnm{{Chen}}, \binits{Q.}},
\bauthor{\bsnm{{Gallagher}}, \binits{P.}},
\bauthor{\bsnm{{Grigis}}, \binits{P.T.}},
\bauthor{\bsnm{{Ji}}, \binits{H.}},
\bauthor{\bsnm{{Liu}}, \binits{W.}},
\bauthor{\bsnm{{Milligan}}, \binits{R.O.}},
\bauthor{\bsnm{{Temmer}}, \binits{M.}}:
\byear{2011},
\batitle{{An Observational Overview of Solar Flares}}.
\bjtitle{\ssr}
\bvolume{159},
\bfpage{19}.
\doiurl{10.1007/s11214-010-9701-8}.
\adsurl{2011SSRv..159...19F}.
\end{barticle}
\endbibitem

\bibitem[\protect\citeauthoryear{{Gruber}
  \textit{et~al.}}{2011}]{2011A&A...533A..61G}
\begin{barticle}
\bauthor{\bsnm{{Gruber}}, \binits{D.}},
\bauthor{\bsnm{{Lachowicz}}, \binits{P.}},
\bauthor{\bsnm{{Bissaldi}}, \binits{E.}},
\bauthor{\bsnm{{Briggs}}, \binits{M.S.}},
\bauthor{\bsnm{{Connaughton}}, \binits{V.}},
\bauthor{\bsnm{{Greiner}}, \binits{J.}},
\bauthor{\bsnm{{van der Horst}}, \binits{A.J.}},
\bauthor{\bsnm{{Kanbach}}, \binits{G.}},
\bauthor{\bsnm{{Rau}}, \binits{A.}},
\bauthor{\bsnm{{Bhat}}, \binits{P.N.}},
\bauthor{\bsnm{{Diehl}}, \binits{R.}},
\bauthor{\bsnm{{von Kienlin}}, \binits{A.}},
\bauthor{\bsnm{{Kippen}}, \binits{R.M.}},
\bauthor{\bsnm{{Meegan}}, \binits{C.A.}},
\bauthor{\bsnm{{Paciesas}}, \binits{W.S.}},
\bauthor{\bsnm{{Preece}}, \binits{R.D.}},
\bauthor{\bsnm{{Wilson-Hodge}}, \binits{C.}}:
\byear{2011},
\batitle{{Quasi-periodic pulsations in solar flares: new clues from the Fermi
  Gamma-Ray Burst Monitor}}.
\bjtitle{\aap}
\bvolume{533},
\bfpage{A61}.
\doiurl{10.1051/0004-6361/201117077}.
\adsurl{2011A\%26A...533A..61G}.
\end{barticle}
\endbibitem

\bibitem[\protect\citeauthoryear{{Hudson} and
  {Warmuth}}{2004}]{2004ApJ...614L..85H}
\begin{barticle}
\bauthor{\bsnm{{Hudson}}, \binits{H.S.}},
\bauthor{\bsnm{{Warmuth}}, \binits{A.}}:
\byear{2004},
\batitle{{Coronal Loop Oscillations and Flare Shock Waves}}.
\bjtitle{\apjl}
\bvolume{614},
\bfpage{L85}.
\doiurl{10.1086/425314}.
\adsurl{2004ApJ...614L..85H}.
\end{barticle}
\endbibitem

\bibitem[\protect\citeauthoryear{{Hudson}
  \textit{et~al.}}{1994}]{1994ApJ...422L..25H}
\begin{barticle}
\bauthor{\bsnm{{Hudson}}, \binits{H.S.}},
\bauthor{\bsnm{{Strong}}, \binits{K.T.}},
\bauthor{\bsnm{{Dennis}}, \binits{B.R.}},
\bauthor{\bsnm{{Zarro}}, \binits{D.}},
\bauthor{\bsnm{{Inda}}, \binits{M.}},
\bauthor{\bsnm{{Kosugi}}, \binits{T.}},
\bauthor{\bsnm{{Sakao}}, \binits{T.}}:
\byear{1994},
\batitle{{Impulsive behavior in solar soft X-radiation}}.
\bjtitle{\apjl}
\bvolume{422},
\bfpage{L25}.
\doiurl{10.1086/187203}.
\adsurl{1994ApJ...422L..25H}.
\end{barticle}
\endbibitem

\bibitem[\protect\citeauthoryear{{Inglis} and
  {Nakariakov}}{2009}]{InglisNakariakov:2009}
\begin{barticle}
\bauthor{\bsnm{{Inglis}}, \binits{A.R.}},
\bauthor{\bsnm{{Nakariakov}}, \binits{V.M.}}:
\byear{2009},
\batitle{{A multi-periodic oscillatory event in a solar flare}}.
\bjtitle{\aap}
\bvolume{493},
\bfpage{259}.
\doiurl{10.1051/0004-6361:200810473}.
\adsurl{2009A\%26A...493..259I}.
\end{barticle}
\endbibitem

\bibitem[\protect\citeauthoryear{{Inglis}, {Ireland}, and
  {Dominique}}{2015}]{InglisIrelandDominique:2015}
\begin{barticle}
\bauthor{\bsnm{{Inglis}}, \binits{A.R.}},
\bauthor{\bsnm{{Ireland}}, \binits{J.}},
\bauthor{\bsnm{{Dominique}}, \binits{M.}}:
\byear{2015},
\batitle{{Quasi-periodic Pulsations in Solar and Stellar Flares: Re-evaluating
  their Nature in the Context of Power-law Flare Fourier Spectra}}.
\bjtitle{\apj}
\bvolume{798},
\bfpage{108}.
\doiurl{10.1088/0004-637X/798/2/108}.
\adsurl{2015ApJ...798..108I}.
\end{barticle}
\endbibitem

\bibitem[\protect\citeauthoryear{{Inglis}, {Nakariakov}, and
  {Melnikov}}{2008}]{2008A&A...487.1147I}
\begin{barticle}
\bauthor{\bsnm{{Inglis}}, \binits{A.R.}},
\bauthor{\bsnm{{Nakariakov}}, \binits{V.M.}},
\bauthor{\bsnm{{Melnikov}}, \binits{V.F.}}:
\byear{2008},
\batitle{{Multi-wavelength spatially resolved analysis of quasi-periodic
  pulsations in a solar flare}}.
\bjtitle{\aap}
\bvolume{487},
\bfpage{1147}.
\doiurl{10.1051/0004-6361:20079323}.
\adsurl{2008A\%26A...487.1147I}.
\end{barticle}
\endbibitem

\bibitem[\protect\citeauthoryear{{Ireland}, {McAteer}, and
  {Inglis}}{2015}]{2015ApJ...798....1I}
\begin{barticle}
\bauthor{\bsnm{{Ireland}}, \binits{J.}},
\bauthor{\bsnm{{McAteer}}, \binits{R.T.J.}},
\bauthor{\bsnm{{Inglis}}, \binits{A.R.}}:
\byear{2015},
\batitle{{Coronal Fourier Power Spectra: Implications for Coronal Seismology
  and Coronal Heating}}.
\bjtitle{\apj}
\bvolume{798},
\bfpage{1}.
\doiurl{10.1088/0004-637X/798/1/1}.
\adsurl{2015ApJ...798....1I}.
\end{barticle}
\endbibitem

\bibitem[\protect\citeauthoryear{{Janssens} and
  {White}}{1969}]{1969ApJ...158L.127J}
\begin{barticle}
\bauthor{\bsnm{{Janssens}}, \binits{T.J.}},
\bauthor{\bsnm{{White}}, \binits{K.P.} \bsuffix{III}}:
\byear{1969},
\batitle{{Microwave Pulse Trains Observed Before and during a Solar Flare}}.
\bjtitle{\apjl}
\bvolume{158},
\bfpage{L127}.
\doiurl{10.1086/180447}.
\adsurl{1969ApJ...158L.127J}.
\end{barticle}
\endbibitem

\bibitem[\protect\citeauthoryear{{Kallunki} and
  {Pohjolainen}}{2012}]{2012SoPh..280..491K}
\begin{barticle}
\bauthor{\bsnm{{Kallunki}}, \binits{J.}},
\bauthor{\bsnm{{Pohjolainen}}, \binits{S.}}:
\byear{2012},
\batitle{{Radio Pulsating Structures with Coronal Loop Contraction}}.
\bjtitle{\solphys}
\bvolume{280},
\bfpage{491}.
\doiurl{10.1007/s11207-012-0003-z}.
\adsurl{2012SoPh..280..491K}.
\end{barticle}
\endbibitem

\bibitem[\protect\citeauthoryear{{Kupriyanova}
  \textit{et~al.}}{2010}]{Kupriyanova:2010}
\begin{barticle}
\bauthor{\bsnm{{Kupriyanova}}, \binits{E.G.}},
\bauthor{\bsnm{{Melnikov}}, \binits{V.F.}},
\bauthor{\bsnm{{Nakariakov}}, \binits{V.M.}},
\bauthor{\bsnm{{Shibasaki}}, \binits{K.}}:
\byear{2010},
\batitle{{Types of Microwave Quasi-Periodic Pulsations in Single Flaring
  Loops}}.
\bjtitle{\solphys}
\bvolume{267},
\bfpage{329}.
\doiurl{10.1007/s11207-010-9642-0}.
\adsurl{2010SoPh..267..329K}.
\end{barticle}
\endbibitem

\bibitem[\protect\citeauthoryear{{Mariska}}{2005}]{2005ApJ...620L..67M}
\begin{barticle}
\bauthor{\bsnm{{Mariska}}, \binits{J.T.}}:
\byear{2005},
\batitle{{Observations of Solar Flare Doppler Shift Oscillations with the Bragg
  Crystal Spectrometer on Yohkoh}}.
\bjtitle{\apjl}
\bvolume{620},
\bfpage{L67}.
\doiurl{10.1086/428611}.
\adsurl{2005ApJ...620L..67M}.
\end{barticle}
\endbibitem

\bibitem[\protect\citeauthoryear{{Mariska}}{2006}]{2006ApJ...639..484M}
\begin{barticle}
\bauthor{\bsnm{{Mariska}}, \binits{J.T.}}:
\byear{2006},
\batitle{{Characteristics of Solar Flare Doppler-Shift Oscillations Observed
  with the Bragg Crystal Spectrometer on Yohkoh}}.
\bjtitle{\apj}
\bvolume{639},
\bfpage{484}.
\doiurl{10.1086/499296}.
\adsurl{2006ApJ...639..484M}.
\end{barticle}
\endbibitem

\bibitem[\protect\citeauthoryear{{McTiernan}
  \textit{et~al.}}{1993}]{1993ApJ...416L..91M}
\begin{barticle}
\bauthor{\bsnm{{McTiernan}}, \binits{J.M.}},
\bauthor{\bsnm{{Kane}}, \binits{S.R.}},
\bauthor{\bsnm{{Loran}}, \binits{J.M.}},
\bauthor{\bsnm{{Lemen}}, \binits{J.R.}},
\bauthor{\bsnm{{Acton}}, \binits{L.W.}},
\bauthor{\bsnm{{Hara}}, \binits{H.}},
\bauthor{\bsnm{{Tsuneta}}, \binits{S.}},
\bauthor{\bsnm{{Kosugi}}, \binits{T.}}:
\byear{1993},
\batitle{{Temperature and Density Structure of the 1991 November 2 Flare Obs
  erved by the YOHKOH Soft X-Ray Telescope and Hard X-Ray Telescope}}.
\bjtitle{\apjl}
\bvolume{416},
\bfpage{L91+}.
\doiurl{10.1086/187078}.
\adsurl{1993ApJ...416L..91M}.
\end{barticle}
\endbibitem

\bibitem[\protect\citeauthoryear{{Meegan} \textit{et~al.}}{2009}]{Meegan:2009}
\begin{barticle}
\bauthor{\bsnm{{Meegan}}, \binits{C.}},
\bauthor{\bsnm{{Lichti}}, \binits{G.}},
\bauthor{\bsnm{{Bhat}}, \binits{P.N.}},
\bauthor{\bsnm{{Bissaldi}}, \binits{E.}},
\bauthor{\bsnm{{Briggs}}, \binits{M.S.}},
\bauthor{\bsnm{{Connaughton}}, \binits{V.}},
\bauthor{\bsnm{{Diehl}}, \binits{R.}},
\bauthor{\bsnm{{Fishman}}, \binits{G.}},
\bauthor{\bsnm{{Greiner}}, \binits{J.}},
\bauthor{\bsnm{{Hoover}}, \binits{A.S.}},
\bauthor{\bsnm{{van der Horst}}, \binits{A.J.}},
\bauthor{\bsnm{{von Kienlin}}, \binits{A.}},
\bauthor{\bsnm{{Kippen}}, \binits{R.M.}},
\bauthor{\bsnm{{Kouveliotou}}, \binits{C.}},
\bauthor{\bsnm{{McBreen}}, \binits{S.}},
\bauthor{\bsnm{{Paciesas}}, \binits{W.S.}},
\bauthor{\bsnm{{Preece}}, \binits{R.}},
\bauthor{\bsnm{{Steinle}}, \binits{H.}},
\bauthor{\bsnm{{Wallace}}, \binits{M.S.}},
\bauthor{\bsnm{{Wilson}}, \binits{R.B.}},
\bauthor{\bsnm{{Wilson-Hodge}}, \binits{C.}}:
\byear{2009},
\batitle{{The Fermi Gamma-ray Burst Monitor}}.
\bjtitle{\apj}
\bvolume{702},
\bfpage{791}.
\doiurl{10.1088/0004-637X/702/1/791}.
\adsurl{2009ApJ...702..791M}.
\end{barticle}
\endbibitem

\bibitem[\protect\citeauthoryear{{Melnikov}
  \textit{et~al.}}{2005}]{2005A&A...439..727M}
\begin{barticle}
\bauthor{\bsnm{{Melnikov}}, \binits{V.F.}},
\bauthor{\bsnm{{Reznikova}}, \binits{V.E.}},
\bauthor{\bsnm{{Shibasaki}}, \binits{K.}},
\bauthor{\bsnm{{Nakariakov}}, \binits{V.M.}}:
\byear{2005},
\batitle{{Spatially resolved microwave pulsations of a flare loop}}.
\bjtitle{\aap}
\bvolume{439},
\bfpage{727}.
\doiurl{10.1051/0004-6361:20052774}.
\adsurl{2005A\%26A...439..727M}.
\end{barticle}
\endbibitem

\bibitem[\protect\citeauthoryear{{Mrozek} and
  {Tomczak}}{2004}]{2004A&A...415..377M}
\begin{barticle}
\bauthor{\bsnm{{Mrozek}}, \binits{T.}},
\bauthor{\bsnm{{Tomczak}}, \binits{M.}}:
\byear{2004},
\batitle{{Solar impulsive soft X-ray brightenings and their connection with
  footpoint hard X-ray emission sources}}.
\bjtitle{\aap}
\bvolume{415},
\bfpage{377}.
\doiurl{10.1051/0004-6361:20034598}.
\adsurl{2004A\%26A...415..377M}.
\end{barticle}
\endbibitem

\bibitem[\protect\citeauthoryear{{Nakariakov} and
  {Melnikov}}{2009}]{2009SSRv..149..119N}
\begin{barticle}
\bauthor{\bsnm{{Nakariakov}}, \binits{V.M.}},
\bauthor{\bsnm{{Melnikov}}, \binits{V.F.}}:
\byear{2009},
\batitle{{Quasi-Periodic Pulsations in Solar Flares}}.
\bjtitle{\ssr}
\bvolume{149},
\bfpage{119}.
\doiurl{10.1007/s11214-009-9536-3}.
\adsurl{2009SSRv..149..119N}.
\end{barticle}
\endbibitem

\bibitem[\protect\citeauthoryear{{Nakariakov} and
  {Verwichte}}{2005}]{2005LRSP....2....3N}
\begin{barticle}
\bauthor{\bsnm{{Nakariakov}}, \binits{V.M.}},
\bauthor{\bsnm{{Verwichte}}, \binits{E.}}:
\byear{2005},
\batitle{{Coronal Waves and Oscillations}}.
\bjtitle{Living Rev. Solar Phys.}
\bvolume{2},
\bfpage{3}.
\doiurl{10.12942/lrsp-2005-3}.
\adsurl{2005LRSP....2....3N}.
\end{barticle}
\endbibitem

\bibitem[\protect\citeauthoryear{{Nakariakov} and
  {Zimovets}}{2011}]{2011ApJ...730L..27N}
\begin{barticle}
\bauthor{\bsnm{{Nakariakov}}, \binits{V.M.}},
\bauthor{\bsnm{{Zimovets}}, \binits{I.V.}}:
\byear{2011},
\batitle{{Slow Magnetoacoustic Waves in Two-ribbon Flares}}.
\bjtitle{\apjl}
\bvolume{730},
\bfpage{L27}.
\doiurl{10.1088/2041-8205/730/2/L27}.
\adsurl{2011ApJ...730L..27N}.
\end{barticle}
\endbibitem

\bibitem[\protect\citeauthoryear{{Nakariakov}, {Melnikov}, and
  {Reznikova}}{2003}]{2003A&A...412L...7N}
\begin{barticle}
\bauthor{\bsnm{{Nakariakov}}, \binits{V.M.}},
\bauthor{\bsnm{{Melnikov}}, \binits{V.F.}},
\bauthor{\bsnm{{Reznikova}}, \binits{V.E.}}:
\byear{2003},
\batitle{{Global sausage modes of coronal loops}}.
\bjtitle{\aap}
\bvolume{412},
\bfpage{L7}.
\doiurl{10.1051/0004-6361:20031660}.
\adsurl{2003A\%26A...412L...7N}.
\end{barticle}
\endbibitem

\bibitem[\protect\citeauthoryear{{Neupert}}{1968}]{1968ApJ...153L..59N}
\begin{barticle}
\bauthor{\bsnm{{Neupert}}, \binits{W.M.}}:
\byear{1968},
\batitle{{Comparison of Solar X-Ray Line Emission with Microwave Emission
  during Flares}}.
\bjtitle{\apjl}
\bvolume{153},
\bfpage{L59}.
\doiurl{10.1086/180220}.
\adsurl{1968ApJ...153L..59N}.
\end{barticle}
\endbibitem

\bibitem[\protect\citeauthoryear{{O'Dwyer}
  \textit{et~al.}}{2010}]{2010A&A...521A..21O}
\begin{barticle}
\bauthor{\bsnm{{O'Dwyer}}, \binits{B.}},
\bauthor{\bsnm{{Del Zanna}}, \binits{G.}},
\bauthor{\bsnm{{Mason}}, \binits{H.E.}},
\bauthor{\bsnm{{Weber}}, \binits{M.A.}},
\bauthor{\bsnm{{Tripathi}}, \binits{D.}}:
\byear{2010},
\batitle{{SDO/AIA response to coronal hole, quiet Sun, active region, and flare
  plasma}}.
\bjtitle{\aap}
\bvolume{521},
\bfpage{A21}.
\doiurl{10.1051/0004-6361/201014872}.
\adsurl{2010A\%26A...521A..21O}.
\end{barticle}
\endbibitem

\bibitem[\protect\citeauthoryear{{Paciesas}}{2011}]{2011AIPC.1366..155P}
\begin{bchapter}
\bauthor{\bsnm{{Paciesas}}, \binits{W.S.}}:
\byear{2011},
\bctitle{{Fermi GBM: Instrument Description and Science Highlights}}.
In: \beditor{\bsnm{{Florinski}}, \binits{V.}},
\beditor{\bsnm{{Heerikhuisen}}, \binits{J.}},
\beditor{\bsnm{{Zank}}, \binits{G.P.}},
\beditor{\bsnm{{Gallagher}}, \binits{D.L.}} (eds.)
\bbtitle{American Institute of Physics Conference Series},
\bsertitle{American Institute of Physics Conference Series}
\bseriesno{1366},
\bfpage{155}.
\doiurl{10.1063/1.3625601}.
\adsurl{2011AIPC.1366..155P}.
\end{bchapter}
\endbibitem

\bibitem[\protect\citeauthoryear{{Parks} and
  {Winckler}}{1969}]{1969ApJ...155L.117P}
\begin{barticle}
\bauthor{\bsnm{{Parks}}, \binits{G.K.}},
\bauthor{\bsnm{{Winckler}}, \binits{J.R.}}:
\byear{1969},
\batitle{{Sixteen-Second Periodic Pulsations Observed in the Correlated
  Microwave and Energetic X-Ray Emission from a Solar Flare}}.
\bjtitle{\apjl}
\bvolume{155},
\bfpage{L117}.
\doiurl{10.1086/180315}.
\adsurl{1969ApJ...155L.117P}.
\end{barticle}
\endbibitem

\bibitem[\protect\citeauthoryear{{Reznikova}
  \textit{et~al.}}{2007}]{2007ARep...51..588R}
\begin{barticle}
\bauthor{\bsnm{{Reznikova}}, \binits{V.E.}},
\bauthor{\bsnm{{Melnikov}}, \binits{V.F.}},
\bauthor{\bsnm{{Su}}, \binits{Y.}},
\bauthor{\bsnm{{Huang}}, \binits{G.}}:
\byear{2007},
\batitle{{Pulsations of microwave flaring emission at low and high
  frequencies}}.
\bjtitle{Astronomy Reports}
\bvolume{51},
\bfpage{588}.
\doiurl{10.1134/S1063772907070086}.
\adsurl{2007ARep...51..588R}.
\end{barticle}
\endbibitem

\bibitem[\protect\citeauthoryear{{Schwartz}
  \textit{et~al.}}{2010}]{2010AAS...21640406S}
\begin{bchapter}
\bauthor{\bsnm{{Schwartz}}, \binits{R.A.}},
\bauthor{\bsnm{{Dennis}}, \binits{B.}},
\bauthor{\bsnm{{Tolbert}}, \binits{A.K.}},
\bauthor{\bsnm{{Murphy}}, \binits{R.}},
\bauthor{\bsnm{{Share}}, \binits{G.}},
\bauthor{\bsnm{{Fishman}}, \binits{G.}},
\bauthor{\bsnm{{Briggs}}, \binits{M.}},
\bauthor{\bsnm{{Longo}}, \binits{F.}},
\bauthor{\bsnm{{Diehl}}, \binits{R.}},
\bauthor{\bsnm{{Wijers}}, \binits{R.}}:
\byear{2010},
\bctitle{{Fermi GBM and LAT Solar Flare X Ray and {$\gamma$} Ray
  Observations}}.
In: \bbtitle{American Astronomical Society Meeting Abstracts \#216},
\bsertitle{\baas}
\bseriesno{41},
\bfpage{\#404.06}.
\adsurl{2010AAS...21640406S}.
\end{bchapter}
\endbibitem

\bibitem[\protect\citeauthoryear{{Sim\~oes}, {Graham}, and
  {Fletcher}}{}]{2015Simoes}
\begin{botherref}
\oauthor{\bsnm{{Sim\~oes}}, \binits{P.J.A.}},
\oauthor{\bsnm{{Graham}}, \binits{D.}},
\oauthor{\bsnm{{Fletcher}}, \binits{L.}}:
{Impulsive heating of solar flare ribbons above 10 MK}.
\textit{\solphys}.
\end{botherref}
\endbibitem

\bibitem[\protect\citeauthoryear{{Sim{\~o}es}
  \textit{et~al.}}{2013}]{2013ApJ...777..152S}
\begin{barticle}
\bauthor{\bsnm{{Sim{\~o}es}}, \binits{P.J.A.}},
\bauthor{\bsnm{{Fletcher}}, \binits{L.}},
\bauthor{\bsnm{{Hudson}}, \binits{H.S.}},
\bauthor{\bsnm{{Russell}}, \binits{A.J.B.}}:
\byear{2013},
\batitle{{Implosion of Coronal Loops during the Impulsive Phase of a Solar
  Flare}}.
\bjtitle{\apj}
\bvolume{777},
\bfpage{152}.
\doiurl{10.1088/0004-637X/777/2/152}.
\adsurl{2013ApJ...777..152S}.
\end{barticle}
\endbibitem

\bibitem[\protect\citeauthoryear{{Stepanov}
  \textit{et~al.}}{2004}]{2004AstL...30..480S}
\begin{barticle}
\bauthor{\bsnm{{Stepanov}}, \binits{A.V.}},
\bauthor{\bsnm{{Kopylova}}, \binits{Y.G.}},
\bauthor{\bsnm{{Tsap}}, \binits{Y.T.}},
\bauthor{\bsnm{{Shibasaki}}, \binits{K.}},
\bauthor{\bsnm{{Melnikov}}, \binits{V.F.}},
\bauthor{\bsnm{{Goldvarg}}, \binits{T.B.}}:
\byear{2004},
\batitle{{Pulsations of Microwave Emission and Flare Plasma Diagnostics}}.
\bjtitle{Astron. Lett.}
\bvolume{30},
\bfpage{480}.
\doiurl{10.1134/1.1774400}.
\adsurl{2004AstL...30..480S}.
\end{barticle}
\endbibitem

\bibitem[\protect\citeauthoryear{{Svestka}}{1976}]{1976sofl.book.....S}
\begin{bbook}
\bauthor{\bsnm{{Svestka}}, \binits{Z.}}:
\byear{1976},
\bbtitle{{Solar Flares}},
\bpublisher{Springer},
\blocation{Berlin Heidelberg}.
\adsurl{1976sofl.book.....S}.
\end{bbook}
\endbibitem

\bibitem[\protect\citeauthoryear{{Torrence} and
  {Compo}}{1998}]{TorrenceCompo:1998}
\begin{botherref}
\oauthor{\bsnm{{Torrence}}, \binits{C.}},
\oauthor{\bsnm{{Compo}}, \binits{G.P.}}:
1998,
A practical guide to wavelet analysis.
\textit{Bull. Am. Meteorol. Soc.},
61.
\end{botherref}
\endbibitem

\bibitem[\protect\citeauthoryear{{Vaughan}}{2005}]{Vaughan:2005}
\begin{barticle}
\bauthor{\bsnm{{Vaughan}}, \binits{S.}}:
\byear{2005},
\batitle{{A simple test for periodic signals in red noise}}.
\bjtitle{\aap}
\bvolume{431},
\bfpage{391}.
\doiurl{10.1051/0004-6361:20041453}.
\adsurl{2005A\%26A...431..391V}.
\end{barticle}
\endbibitem

\bibitem[\protect\citeauthoryear{{Vaughan}}{2010}]{Vaughan:2010}
\begin{barticle}
\bauthor{\bsnm{{Vaughan}}, \binits{S.}}:
\byear{2010},
\batitle{{A Bayesian test for periodic signals in red noise}}.
\bjtitle{\mnras}
\bvolume{402},
\bfpage{307}.
\doiurl{10.1111/j.1365-2966.2009.15868.x}.
\adsurl{2010MNRAS.402..307V}.
\end{barticle}
\endbibitem

\bibitem[\protect\citeauthoryear{{White}, {Thomas}, and
  {Schwartz}}{2005}]{2005SoPh..227..231W}
\begin{barticle}
\bauthor{\bsnm{{White}}, \binits{S.M.}},
\bauthor{\bsnm{{Thomas}}, \binits{R.J.}},
\bauthor{\bsnm{{Schwartz}}, \binits{R.A.}}:
\byear{2005},
\batitle{{Updated Expressions for Determining Temperatures and Emission Meas
  ures from Goes Soft X-Ray Measurements}}.
\bjtitle{\solphys}
\bvolume{227},
\bfpage{231}.
\doiurl{10.1007/s11207-005-2445-z}.
\adsurl{2005SoPh..227..231W}.
\end{barticle}
\endbibitem

\end{thebibliography}
%
%
%
%

\end{article} 
\end{document}